\documentclass[pra,superscriptaddress,reprint]{revtex4-1}

\usepackage{amsmath}
\usepackage{graphicx}
\usepackage{bm}
\usepackage{epstopdf}

\newcommand{\be}{\begin{equation}}
\newcommand{\ee}{\end{equation}}
\newcommand{\bea}{\begin{eqnarray}}
\newcommand{\eea}{\end{eqnarray}}
\newcommand{\ket}[1]{\left\vert #1    \right\rangle }
\newcommand{\bra}[1]{\left\langle   #1  \right\vert}
\newcommand{\ave}[1]{\left\langle #1   \right\rangle }

\newcommand{\w}{\omega}
\newcommand{\W}{\Omega}
\newcommand{\unit}[1]{\ensuremath{\, \mathrm{#1}}}

\begin{document}

\title{Double-path dark-state laser cooling in a three-level system}

\author{J. Cerrillo}
\email{cerrillo@tu-berlin.de}
\affiliation{Institut f\"ur Theoretische Physik, Technische Universit\"at Berlin, Hardenbergstrasse 36, D-10623 Berlin, Germany}
\author{A. Retzker}
\affiliation{Racah Institute of Physics, The Hebrew University of Jerusalem, Jerusalem 91904, Givat Ram, Israel}
\author{M. B. Plenio}
\affiliation{Institut f\"ur Theoretische Physik und IQST, Albert-Einstein-Allee 11, Univers\"at Ulm, D-89069 Ulm, Germany}

\begin{abstract}
We present a detailed analysis of a robust and fast laser cooling scheme  [J. Cerrillo {\em et.~al.}, Phys.~Rev.~Lett.~{\bf 104}, 043003 (2010)] on a three level system. A special laser configuration, applicable to trapped ions, atoms or cantilevers, designs a double path quantum interference that eliminates the blue sideband in addition to the carrier transition, thus excluding any heating process involving up to one-phonon interactions. As a consequence, cooling achieves vanishing phonon occupation up to first order in the Lamb-Dicke parameter expansion. Underlying this scheme is a combined action of two cooling schemes which makes the proposal very flexible under constraints of the physical parameters such as laser intensity, detuning or optical access, making it a viable candidate for experimental implementation. Furthermore, it is considerably faster than existing ground state cooling schemes. Its suitability as a cooling scheme for several ions in a trap and 3D cooling is shown.
\end{abstract}

\maketitle

\section{Introduction}

The use of laser cooling schemes \cite*{Chu1998,Cohen-Tannoudji1998,Phillips1998} for the motional degrees of freedom of trapped particles has proven effective and useful in the initialization of experiments within the quantum regime. Whether the particle is free or bound by an external potential dictates a fundamental distinction among different treatments. The idea underlying Doppler cooling \cite{Hansch1975} for free particles is related to sideband cooling \cite{Wineland1975,Wineland1978} for bound particles and similarly dark state cooling for free particles \cite{Aspect1988} has its counterpart for trapped ions \cite{Dum1994}. At present, sideband cooling is the method of choice for ground state cooling of trapped ions. By addressing the red motional sideband of an electronic transition with laser light detuned by the value of the trap frequency $\nu$, phonons are scattered away with every spontaneous emission. This requires a transition linewidth $\Gamma$ that is small enough for the sidebands to be resolved, $\Gamma\ll\nu$. Off-resonant heating processes (primarily carrier transition excitation but also blue sideband heating) limit its performance both in terms of cooling rate and final temperature. Its cooling rate is determined by $\Gamma$ and the coupling strength of the laser light to the electronic levels, corresponding to the Rabi frequency $\Omega$ times the Lamb-Dicke parameter $\eta$. The minimum reachable phonon number is limited by $\left(\frac{\Gamma}{4\nu}\right)^{2}+O(\eta^2)$ in the case of very weak driving.

The limitations of sideband cooling can be overcome by involving a third electronic level. On the one hand, this may be used to adjust the effective linewitdh of the transition \cite{Marzoli1994, Monroe1995}. Alternatively, the destructive interference that generates electromagnetically induced transparency (EIT) \cite{Fleischhauer2005} may be used to design dark state cooling schemes \cite{Morigi2000,Roos2000, Morigi2003} that overcome the detrimental effect of the heating associated to the carrier transition. In a three level lambda system, a Raman coupling dresses the electronic states giving rise to one dark state and two bright states. By adjusting the detuning $\Delta$ and laser intensity $\Omega$ correctly, an effective coupling of the dark state with the narrowest bright state can be achieved with a detuning equal to the trap frequency \cite{Morigi2000, Roos2000, Morigi2003}. The use of a dark state ensures the cancellation of the carrier transition, and the detuning adjustment enhances the red sideband transition with respect to blue sideband excitations. All in all, final ocupation numbers proportional to $\left(\frac{\Gamma}{4 \Delta}\right)^2+O(\eta^2) $ can be achieved, while the rate is upper bounded by $\frac{\eta^2\Omega^2}{2\Gamma}$, thus beating sideband cooling for large detunings.

The Stark shift (SSh) cooling method \cite{Retzker2007} is another instance of a dark state cooling scheme for a three level lambda system. In this case the focus resides upon the lowest lying states. A microwave coupling imparts a finite but small momentum on the transition so that a quantum gate \cite{Jonathan2000} can be tailored to involve the mechanical and electronic degrees of freedom. A simple Raman coupling is designed to introduce broadening in such a way that more energetic mechanical states are coupled to a dissipative electronic level. If tuned to the value of the trap frequency, a red sideband coupling is favored and the carrier transition is canceled. Higher values of laser intensities can be applied with this proposal, allowing for an effectively faster operation of the scheme.

The limiting factor on both EIT and SSh coolings is the heating associated to the off-resonant blue sideband which, after the carrier transition, is the only heating process left up to first order phonon processes. For large $\Gamma$ or small $\Delta$ it is not possible to neglect its effect and cooling efficiency diminishes. With the cancellation of the blue sideband, perfect ground state cooling up to first order in the Lamb-Dicke parameter is expected. Furthermore, the EIT cooling rate upper bound $\frac{\eta^2\Omega^2}{2\Gamma}$ may be saturated regardless of the detuning $\Delta$. The first laser cooling proposal that cancels both carrier and blue sideband processes was \citep{Evers2004}. In order to achieve this objective, it introduces a fourth state in a tripod configuration that couples to the excited state, in such form as to establish a blue detuning with respect to the Raman configuration of exactly one trap frequency. The main effect of this extra coupling on the cooling spectrum is the appearance of a zero at exactly the location of the blue sideband process, thus cancelling it. This is in addition to the zero at vanishing frequency already present in single EIT cooling. Cancellation of the blue sideband can also be achieved in another manner that does not involve a fourth electronic level. As first proposed in \cite{Cerrillo2010}, the judicious combination of two laser cooling schemes on the same three level system shifts the already existing zero of single EIT cooling to the location of the blue sideband. In order to achieve this shift, it is necessary that the first order Lamb-Dicke term couples the dark state to two other states, rather than one. This enables a double path quantum interference which characterizes the scheme and introduces a degree of freedom that is exploited to cancel the blue sideband, as opposed to the zeroth order coupling to an additional level of \citep{Evers2004}. To the same class of laser cooling schemes belongs the experimental implementation in \cite{Scharnhorst2017}, where the double path interference is used to address the cooling of several vibrational modes simultaneously. It is also possible to exploit this aspect in \citep{Cerrillo2010}.

In this paper we set out in detail the mechanism presented in \cite{Cerrillo2010}, with particular emphasis on possible experimental implementations. The paper is organized as follows. In section \ref{sec:model} a model of the laser cooling scheme is presented together with the Lamb-Dicke expansion that is used. In section \ref{sec:theory}, the derivation of rate and final temperature limits within the perturbative regime is presented, where the coherent superposition of the effect of both laser schemes becomes apparent. An analysis of the limitations and corrections to the perturbative reasults is found in section \ref{sec:correction}. In section \ref{sec:implementation}, two possible experimental implementations are proposed, and the possibility of cooling the motion on more than one mode or axis with the same scheme is presented in section \ref{sec:multiaxial}. We finalize de discussion with some concluding remarks.

\section{Model and Lamb-Dicke expansion\label{sec:model}}

\begin{figure}
\begin{center}
\includegraphics[width=0.8\columnwidth]{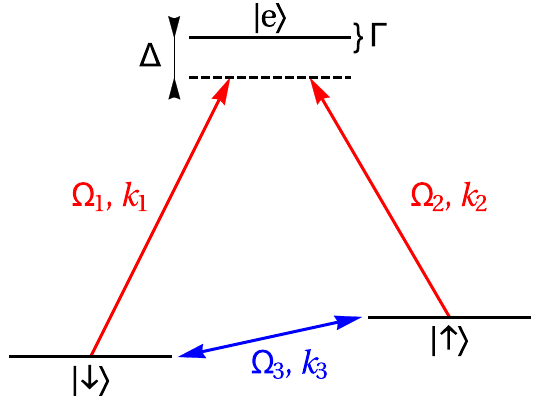}
\end{center}
\caption[Double dark state cooling scheme layout.]{We regard a 3-level electronic system that consists of a ground state $\ket\downarrow$, a metastable state
$\ket\uparrow$ and an excited state $\ket e$ which dissipates
energy at rate $\Gamma$. The lower levels are coupled to $\ket e$
by a pair of Raman beams under a detuning $\Delta$. The beams carry a Rabi frequency $\Omega_1$ and $\W_2$ respectively and their wavevector projections on the cooling axis are $k_1$ and $k_2$. Additionally, $\ket{\uparrow}$ and $\ket{\downarrow}$ are directly coupled by a beam of Rabi
frequency $\Omega_3$ and wavevector projection $k_3$.}
\label{levels}
\end{figure}

This proposal is designed for a 3-level $\Lambda$-system of mass $m$ which is trapped in a harmonic well of frequency $\nu$. This is an accurate model for an ion in an electromagnetic trap or an atom in a deep dipole trap. We consider a metastable state $\ket\uparrow$, a ground state $\ket\downarrow$ and a dissipative excited state $\ket e$, with spontaneous decay rate $\Gamma$. The three levels are coupled by means of an electric dipole interaction with running waves as shown in figure \ref{levels} and expressed in the Hamiltonian
\be\begin{split}
H&= \hbar\nu b^\dagger b + \hbar\omega_e \ket{e} \bra{e} + \hbar\omega_{\uparrow} \ket{\uparrow} \bra{\uparrow}+ \hbar\omega_{\downarrow} \ket{\downarrow} \bra{\downarrow} \\
&+\hbar\Omega_1 \left( \ket{e} \bra{\downarrow} + H.c. \right) \cos\left( \omega_1 t- k_1 x - \phi_1\right) \\
&+\hbar\Omega_2 \left( \ket{e} \bra{\uparrow} + H.c. \right) \cos\left( \omega_2 t- k_2 x - \phi_2\right) \\
&+\hbar\Omega_3 \left( \ket{\uparrow} \bra{\downarrow} + H.c. \right) \cos\left( \omega_3 t - k_3 x - \phi_3\right),
\end{split}\ee
where $\Omega_j$, $\omega_j$, $k_j$ and $\phi_j$ with $j\in\{1,2,3\}$ are respectively the Rabi frequencies, the laser frequencies, the wavevector projections on the cooling axis and the initial phases of the respective laser couplings, and $\hbar\omega_{e,\uparrow,\downarrow}$ are the energies of the respective levels. The Raman condition establishes an overall detuning $\Delta=\omega_e-\omega_{\downarrow}-\omega_1=\omega_e-\omega_{\uparrow}-\omega_2$ and the third laser beam is in resonance $\omega_{\uparrow}-\omega_{\downarrow}=\omega_3$. It is useful to express each wavevector projection in terms of its corresponding Lamb-Dicke parameter following the definition $\eta_j=k_j x_0$, where $x_0=\sqrt{\frac{\hbar}{2 m\nu}}$ is the zero point motion of the oscillator.

The effect of the spontaneous emission is described using a master equation of the form
\be
\frac{d\rho}{dt}=\frac{-i}{\hbar}\left[H,\rho\right]+\mathcal{L}^d(\rho)=\mathcal{L}(\rho),
\label{master}
\ee
where $\rho$ is the state of the system involving both the internal and the external degrees of freedom. The superoperator $\mathcal{L}^d$ is a Lindbladian for the two dissipative channels
\be
\mathcal{L}^d (\rho)= \sum_{i=\downarrow ,\uparrow} \gamma_{i} \left(2  \ket{i}\bra{e}\overline{\rho_{e,i}} \ket{e}\bra{i} -
\rho \ket{e}\bra{e}  - \ket{e}\bra{e}  \rho \right),
\label{diss}
\ee
where the spontaneous decay rates add up to the total rate $\gamma_{\downarrow}+\gamma_\uparrow=\Gamma$  and
\be\overline{\rho_{e,i}} = \frac{1}{2}\int_{-1}^1 ds W(s)e^{ik_{e,i}xs} \rho e^{-ik_{e,i}xs}\ee
accounts for the momentum transfer of $\hbar k_{e,i}$ in the event of a photon emission due to an electronic decay from level $\ket{e}$ to level $\ket{i}$. $W(s)=\frac 3 4 (1+s^2)$ is the angular distribution for a spontaneous emission of a dipole transition.

In this section we consider a perturbative expansion in the Lamb-Dicke parameters, which requires the condition  $\eta_j\sqrt{2\ave{n}+1}\ll1$, with $j\in\{1,2,3\}$ and $\ave{n}$ the average occupation number of the vibrational mode of the particle. In physical terms, this implies that the recoil energy gained in each photon emission is much smaller than the energy necessary to excite a motional quantum, therefore processes involving phonon creation or annihilation are realized with small probability. An expansion of the Hamiltonian up to second order phonon processes is hence justified.
Additionally, the laser frequencies involved are large compared to all other timescales of the system, which justifies the use of a rotating wave approximation. In the interaction picture with respect to the laser frequencies, the rotating Hamiltonian $H'$ can be split into the following three terms
\be
H'=H_{m}+H_{e}+V,
\label{Ham}
\ee
with the Hamiltonian for the motional degrees of freedom
\be\begin{split}
H_{m}&= \hbar\nu b^\dagger b,
\end{split}\ee
the Hamiltonian involving only the electronic degrees of freedom
\be\begin{split}
H_{e}&=\hbar\Delta \ket{e} \bra{e} +\frac{\hbar}{2} \Big( \Omega_1 e^{i\phi_1}\ket{e} \bra{\downarrow}\\
& + \Omega_2 e^{i\phi_2} \ket{e} \bra{\uparrow} + \Omega_3 e^{i\phi_3} \ket{\uparrow} \bra{\downarrow} + H.c. \Big),
\end{split}\ee
and the linear coupling between the motional and electronic degrees of freedom
\be\begin{split}
V&=\frac{\hbar}{2}\left( b^\dagger + b\right)\Big(i\eta_1\Omega_1 e^{i\phi_1} \ket{e} \bra{\downarrow} + i\eta_2\Omega_2 e^{i\phi_2} \ket{e} \bra{\uparrow} \\
&+i\eta_3\Omega_3 e^{i\phi_3} \ket{\uparrow} \bra{\downarrow} + H.c. \Big).
\end{split}\ee
This Hamiltonian can be regarded as an interpolation between two cooling schemes. For $\Omega_3=0$, the EIT-cooling Hamiltonian \cite{Morigi2000} is recovered, whereas the limit $\eta_1=\eta_2=0$ corresponds to the Stark-shift-cooling \cite{Retzker2007}.

In the limit $\eta_j=0$, the electron will not scatter any light as long as it is trapped in an eigenstate of $H_e$ that contains no overlap with the excited state. Such a dark state exists always in the EIT limit $\Omega_3=0$, whereas for $\Omega_3\neq 0$ this occurs under the conditions
\be\begin{split}
\phi_3&=\phi_1-\phi_2 + k \pi,\quad k \in Z; \\
\Omega_1&=\Omega_2;
\end{split}\ee
i.~e., under coinciding Rabi frequencies of the Raman pair and locking of their beating phase to the phase of the third beam or its counterphase. These conditions are commonly accessible in experimental realizations and result in the formation of the dark state $\ket{-}=\frac{1}{\sqrt{2}}\left(\ket{\downarrow}-e^{i\phi_3}\ket{\uparrow}\right)$. Together with the bright state $\ket{+}=\frac{e^{-i\phi_1}}{\sqrt{2}}\left(\ket{\downarrow}+e^{i\phi_3}\ket{\uparrow}\right)$, orthogonal to $\ket{-}$, the Hamiltonian $H_e$ takes the simplified form
\be\begin{split}
H_{e}&=\hbar\Delta \ket{e} \bra{e} +\frac{\hbar\Omega_3}{2} \left(\ket{+} \bra{+} - \ket{-} \bra{-} \right)\\
&+\frac{\hbar\Omega}{2} \Big( \ket{e} \bra{+} + H.c. \Big),
\label{eq:He}
\end{split}\ee
where $\Omega\equiv\sqrt{2}\Omega_1$. It becomes apparent that, just as in EIT-cooling, the dark state is decoupled from the laser. Nevertheless, it experiences an additional Stark-shift that is not present in EIT-cooling and that energetically separates it from the bright state $\ket{+}$. As will become apparent below, this Stark-shift is central for the creation of a second dark-state effect in the system. As for the coupling to the motional degrees of freedom, the Hamiltonian $V$ becomes
\be\begin{split}
V&=\hbar \sigma_\eta \left( b^\dagger + b\right) =\frac{\hbar}{2}\left( b^\dagger + b\right)\Big(i\bar\eta\Omega \ket{e} \bra{+}\\
& + i\eta\Omega e^{i\phi_1} \ket{e} \bra{-} +i\eta_3\Omega_3 e^{i\phi_1} \ket{+} \bra{-} + H.c. \Big),
\end{split}\ee
with $\bar\eta=\frac{\eta_{1}+\eta_{2}}{2}$ and $\eta=\frac{\eta_{1}-\eta_{2}}{2}$. The operator $\sigma_\eta$ couples the dark state both to the excited and the bright states, which constitutes a further control mechanism over the second dark-state effect.

Regarding spontaneous emission, the leading order of $\mathcal{L}(\rho)$ becomes
\be\label{eq:L0}\begin{split}
\mathcal{L}_0  &(\rho)=-i[H_m+H_e,\rho]\\
&+\sum_{i=\downarrow ,\uparrow}   \gamma_{i} \left(2  \ket{i}\bra{e} \rho \ket{e}\bra{i} -
\rho \ket{e}\bra{e}  - \ket{e}\bra{e}  \rho \right),
\end{split}\ee
and it establishes the dark state $\ket{-}$ as the steady state for the electronic degrees of freedom.

\section{Cooling rate and final temperature\label{sec:theory}}

Within the limit where the electronic dynamics is much faster than the motional degrees of freedom and they are both weakly coupled to each other, adiabatic elimination of the electronic degrees of freedom to second perturbative order is justified \cite{Marzoli1994}. In this regime, the dynamics of the expected value $\ave{n}(t)$ of the number operator of the vibrational mode follows the equation
\be
\frac{d}{dt}\ave{n}(t)=-(A_--A_+)\ave{n}(t)+A_+,
\ee
where $A_+$ and $A_-$ are the heating and cooling rates respectively, defined as
\be
A_\pm=2Re\left[S(\mp\nu)\right]\label{rates}
\ee
in terms of the Fourier transform of the stationary correlation function of $\sigma_\eta$
\be
S(\w)=\int_0^\infty d\tau e^{i\w\tau} \ave{\sigma_{\eta}(\tau)\sigma_{\eta}},\label{spectrum}
\ee
where the average is over the electronic steady state $\ket{-}$ and $\sigma_\eta$ is the electronic operator of the first order Hamiltonian
\be
\begin{split}
\sigma_\eta =\frac{1}{2}\Big(&i\bar\eta\Omega \ket{e} \bra{+}+ i\eta\Omega e^{i\phi_1} \ket{e} \bra{-} \\
& +i\eta_3\Omega_3 e^{i\phi_1} \ket{+} \bra{-} + H.c. \Big).
\end{split}\ee
Therefore, the motional degrees of freedom cool at the rate
\be
R=A_--A_+,
\ee
to a final occupation
\be
\ave n_f=\frac{A_+}{A_--A_+}.
\ee
In view of Eq.(\ref{rates}), $A_-$ can be interpreted as the ability of the three level system to absorb red $\nu$-detuned light while absorbing a phonon from its motional degrees of freedom. The rate $A_+$ corresponds to the phonon emission rate due to the blue $\nu$-detuned light.

\begin{figure}
\begin{center}
  \includegraphics[trim=0mm 0mm 0mm 30mm, clip, width=\columnwidth]{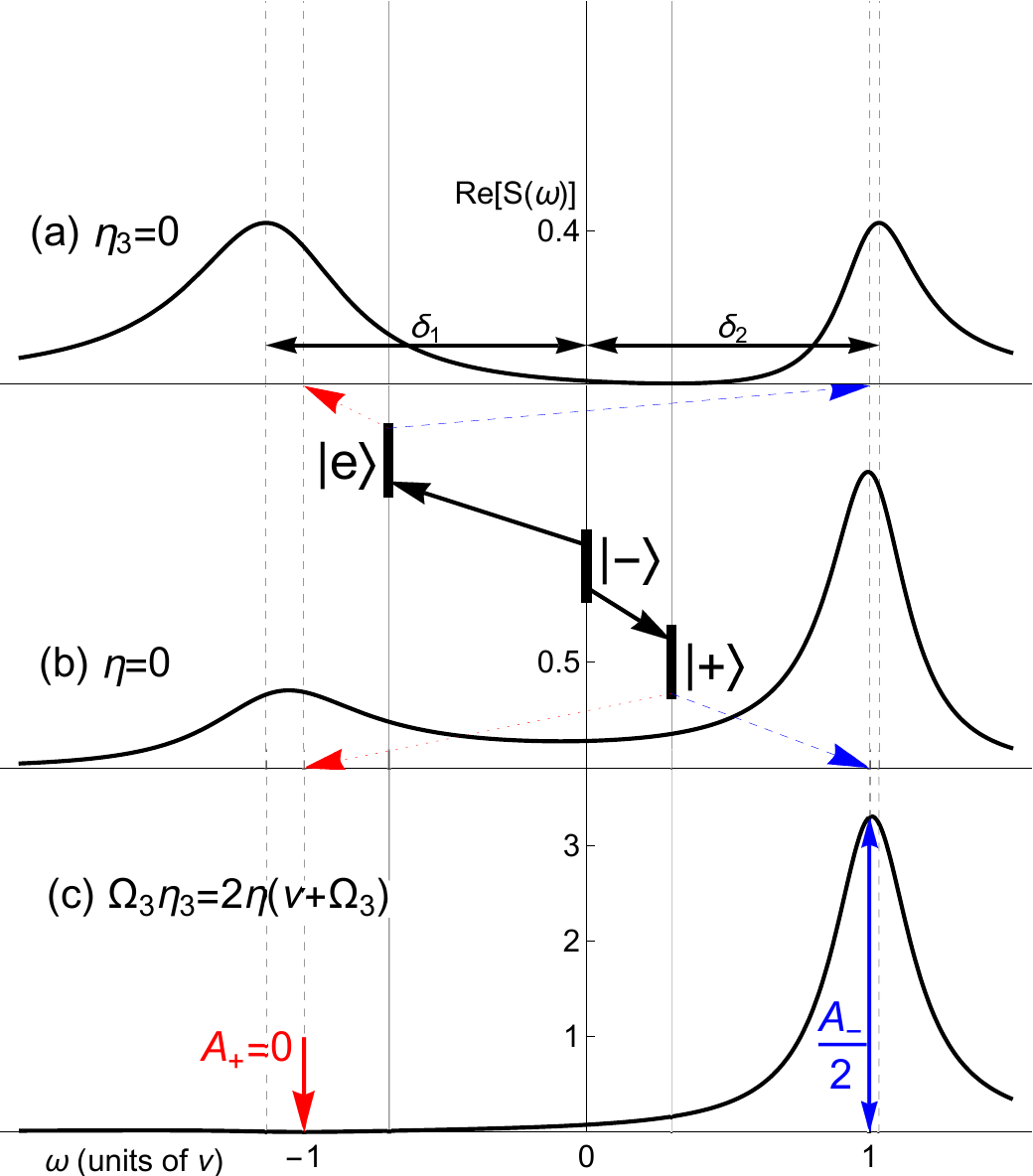}
\end{center}
   \caption{Representation of the absorption spectrum (c) of Eq.(\ref{realspec}) as a superposition of two components associated respectively to (a) the excited state $\ket e$ and (b) the bright state  $\ket +$. Spectrum (a) is characteristic of the EIT scheme and is the result of setting $\eta_3=0$ in  Eq.(\ref{realspec}). It contains two maxima, located at the eigenfrequencies $\delta_1$ and $\delta_2$ of the Stark shifted states $\ket{D_1}$ and  $\ket{D_2}$ respectively. Additionally, it vanishes at the eigenfrequency of the bright state $\ket{+}$. Spectrum (b) arises in the context of SSh cooling and results from Eq.(\ref{realspec}) for $\eta=0$. Its maxima do not coincide with the stark shifted states. Spectrum (c) contains both contributions under condition Eq.(\ref{RobustCond}). Its evaluation at the motional frequency provides half of the cooling rate $A_-$. The heating rate $A_+$ vanishes under the opposing contributions of (a) and (b). Parameters: $\Gamma=0.5\nu$, $\Delta=-0.55\nu$, $\W=2.05\nu$, $\W_3=0.3\nu$. In (a) and (c), $\eta=0.45$, in (b) $\eta_3=4$ (Lamb-Dicke parameters have been exagerated for graphic clarity).}
\label{fig:interference}
\end{figure}

The real part of the spectrum Eq.(\ref{spectrum}) becomes within the quantum regression theorem (see appendix \ref{sec:spectrumreal})
\be
Re\left[S(\w)\right]=\frac{\Gamma\left|\bra{\eta} M\left(\w\right)^{-1}\ket{e}\right|^2}{1+\Gamma^2\left|\bra{e}M\left(\w\right)^{-1}\ket{e}\right|^2}.\label{realspec}
\ee
with $\ket{\eta}=\frac{1}{2}\left(\eta\Omega\ket{e} + \eta_3\Omega_3\ket{+}\right)$ and
\be
M\left(\w\right)=\left(\begin{array}{cc}
\w-\Delta - \frac{\Omega_3}{2}  & -\frac{\Omega}{2}\\
-\frac{\Omega}{2} & \w- \Omega_3
\end{array}\right),\label{MatrixM}
\ee
for $\bra{e}=(1,0)$ and $\bra{+}=(0,1)$. The cooling effect of the scheme is therefore the result of the coherent superposition of two spectra as determined by $\ket{\eta}$.  As shown in Fig.(\ref{fig:interference}), on the one hand, the coupling to the excited state introduces the characteristic EIT cooling spectrum (Fig.\ref{fig:interference}a); on the other hand, the coupling to the bright state introduces the Stark-shift one (Fig.\ref{fig:interference}b) and the superposition of both produces the absorption spectrum for the present scheme (Fig.\ref{fig:interference}c). An explicit calculation of this superposition yields
\begin{equation}
Re\left[S(\w)\right]=\frac{1}{16}\frac{\Gamma\Omega^2\left[2\eta(\w - \Omega_3)+\eta_3\Omega_3\right]^2}{\left(\w-\delta_1\right)^2\left(\w-\delta_2\right)^2 + \Gamma^2(\w - \W_3)^2},\label{eq:spectrum}
\end{equation}
where we use the eigenfrequencies (shifted by $-\frac{\W_3}{2}$) $\delta_1$ and $\delta_2$ of the zeroth order electronic Hamiltonian $H_{e}$ of Eq.(\ref{eq:He}) 
\be
\delta_{1,2}=\frac{\Delta+\frac{3}{2}\W_3\mp\sqrt{\left(\Delta-\frac{\W_{3}}{2}\right)^{2}+\Omega^{2}}}{2}.
\ee
corresponding to eigenstates  $\ket{D_1}$ and $\ket{D_2}$, defined by
\be\begin{split}
\label{eq:dressed}
\ket{+}&=\cos\phi\ket{D_1}+\sin\phi\ket{D_2},\\
\ket{e}&=\sin\phi\ket{D_1}-\cos\phi\ket{D_2},
\end{split}\ee
where
\be
\tan\phi=-\frac{\Delta-\frac{\W_{3}}{2}}{\Omega}+\sqrt{\left(\frac{\Delta-\frac{\W_{3}}{2}}{ \Omega}\right)^{2}+1},
\ee
which defines the mixing angle between the excited and the bright states.

The spectrum Eq.(\ref{eq:spectrum}) contains several features that facilitate the design of an efficient cooling scheme. The most relevant one is a frequency $\w_0$ of vanishing absorption determined by $2\eta(\w_0 - \Omega_3)=-\eta_3\Omega_3$. As explicited in \cite{Cerrillo2010}, it can be adjusted in terms of the value of the Lamb-Dicke parameters $\eta$, $\eta_3$ and the Rabi frequency $\W_3$ such that the blue sideband absorption is suppressed by setting $\w_0=-\nu$. This establishes the condition
\be
\frac{\eta_3}{2\eta}=\frac{\nu}{\W_3}+1,
\label{RobustCond}
\ee
where Eq.(\ref{eq:spectrum}) takes the simple form
\begin{equation}
Re\left[S(\w)\right]=\frac{\Omega^2\eta^2}{4}\frac{\Gamma(\w + \nu)^2}{\left(\w-\delta_1\right)^2\left(\w-\delta_2\right)^2 + \Gamma^2(\w - \W_3)^2 },
\label{eq:Robustspectraldensity}
\end{equation}
from where $A_+=0$ follows. The elimination of the blue sideband contribution implies that, within second order perturbative theory, all heating mechanisms vanish and the ground state is reached identically, so that
\be
\ave n_f=0.
\label{non}
\ee
Additionally, the cooling rate is now simply given by $R=A_-$. Several strategies may be followed for its optimization, and we provide a number of them:

1. The {\em Stark-shift condition}
\be
\W_3=\nu,
\label{SShCond}
\ee
discussed in \cite{Retzker2007}, such that a cooling rate of the form
\be
R_{max}=\Gamma  \frac{\eta_3^2\W_3}{\W^2},
\label{eq:RateLD1}
\ee
may be achieved. It is proportional to the trap frequency, with a constant provided by the ratio between the rate of operation of the Stark-shift gate \cite{Jonathan2000} and the effective decay rate of the bright state $\frac{\W^2}{\Gamma}$. A decrease of this effective decay rate is associated to an increase of the cooling rate, although this prediction breaks down in the limit of vanishing $\W$, which is beyond the region of validity of the adiabatic elimination. 

2. The {\em EIT condition}. Alternatively, one may consider 
\be
\W_3=-\nu,
\label{EITCond}
\ee
that corresponds to $\eta_3=0$ following Eq.(\ref{RobustCond}). In this case, the spectral maxima coincide with the eigenfrequencies $\delta_{1,2}$. The condition for maximum rate is then achieved by either $\delta_{1,2}=\nu$ 
and it corresponds to a rate
\be
R_{max}= \frac{\eta^2 \Omega^{2}}{2\Gamma},
\label{eq:RateLD2}
\ee
which saturates the upper bound associated to EIT cooling in the limit of large blue detuning.

3. The {\em robust or multimode condition} consists in adjusting the values of the parameters $\W$, $\W_3$ and $\Delta$ such that the eigenfrequencies of both dressed states $\ket{D_{1,2}}$ Eq.(\ref{eq:dressed}) are larger than that of $\ket -$. This diminishes the negative frequency part of the absorption spectrum while enhancing the positive frequency part. This alternative protects the cooling scheme against intensity or frequency fluctuations and makes it especially suitable for multimode cooling.

In conclusion, there are a number of aspects that make this proposal superior to EIT or SSh cooling schemes and comparable to double EIT proposals, even though just three electronic levels are involved in the design. To begin with, vanishing population is predicted within the Lamb-Dicke regime. This also implies that larger cooling rates may be reached due to the suppression of the heating term. Additionally, the rates may be optimized with independence of the value of the detuning $\Delta$. This is an important aspect that is not available in EIT or SSh cooling where, by construction, one of the dressed states lies on the positive frequency spectrum, while the other one is on the negative frequency part. This inherent limitation imposes the necessity to use a large blue detuning so as to avoid the presence of a resonance in the vicinity of the blue sideband. If the dark state is missed due to experimental imperfections, the cooling scheme becomes a blue detuned Doppler setting, which actually heats the system. Conversely, red detunings can be used in the present proposal for a double dark state scheme, so that failure to match the dark state yields a Doppler cooling scheme with correctly red detuned light, which protects the cooling effect albeit at a lower efficiency rate.

\section{Corrections to the Lamb-Dicke approximation \label{sec:correction}}

The predictions of vanishing phonon number Eq.(\ref{non}) and unlimited rates Eq.(\ref{eq:RateLD1}) and Eq.(\ref{eq:RateLD2}) are only valid in the regime of small Lamb-Dicke parameter and fast electronic dynamics. Beyond these regimes, the perturbative expansion and the adiabatic elimination procedures fail. First order corrections to these predictions are put forward in this section.

\subsection{First order correction to the steady state}

The resulting vanishing population at long times Eq.(\ref{non}) implies a pure steady state of the form \be
\ket{\Psi}_f=\ket{-}\ket{0}.
\ee
Beyond this result, a first correction may be derived in terms of a perturbative approach involving the full master equation Eq.(\ref{master}). Given the form of the dissipator term Eq.(\ref{diss}), a state with no overlap with the excited state $\ket{e}$ that commutes with the full Hamiltonian Eq.(\ref{Ham}) fulfills the conditions of stationarity. To zeroth order, a state fulfilling those conditions is the tensor product of the dark state and an arbitrary vibrational Fock state
\be
\ket{\Psi}^{(0)}_f=\ket{-}\otimes\ket{n},
\label{zerothorder}
\ee
with eigenvalue $\lambda=\hbar\nu n -\frac{\hbar\W_3}{2}$. The same conditions to first order have the form
\be
\begin{split}
\left(H_m+H_e\right)\ket{\Psi}^{(1)}_{f}+V\ket{\Psi}^{(0)}_{f}&=\lambda\ket{\Psi}^{(1)}_{f},\\
\ave{e|\Psi}_{f}^{(1)}&=0.
\end{split}
\ee
Projection of the first equation with $\bra{e}$, $\bra{+}$ and $\bra{-}$ provides 
\be
\begin{split}
\ave{+|\Psi}_{f}^{(1)}&=-i\eta e^{i\phi_1}\bra{-}b^\dagger+b\ket{\Psi}_f^{(0)},\\
\bra{+}H_m+\frac{\hbar\W_3}{2}-\lambda\ket{\Psi}_f^{(1)}&=-i\frac{\hbar\eta_3\W_3}{2}e^{i\phi_1}\bra{-}b^\dagger+b\ket{\Psi}_f^{(0)},\\
\bra{-}H_m-\frac{\hbar\W_3}{2}-\lambda\ket{\Psi}_f^{(1)}&=0,
\end{split}
\ee
which establishes $n=0$ in Eq.(\ref{zerothorder}) and additionally
\be
\ket{\Psi}_{f}^{(1)}=-i \eta\ket{+}\otimes\ket{1},
\ee
under the condition Eq.(\ref{RobustCond}).

Therefore, a better approximation for the final occupation number $\ave n_f$ is rather $\eta^2$. Nevertheless, the steady state remains pure for the first two orders in the perturbative expansion and does not contribute to the final temperature. 

\subsection{Breakdown of cooling rate predictions}

\begin{figure}
\begin{center}
\includegraphics[width=\columnwidth]{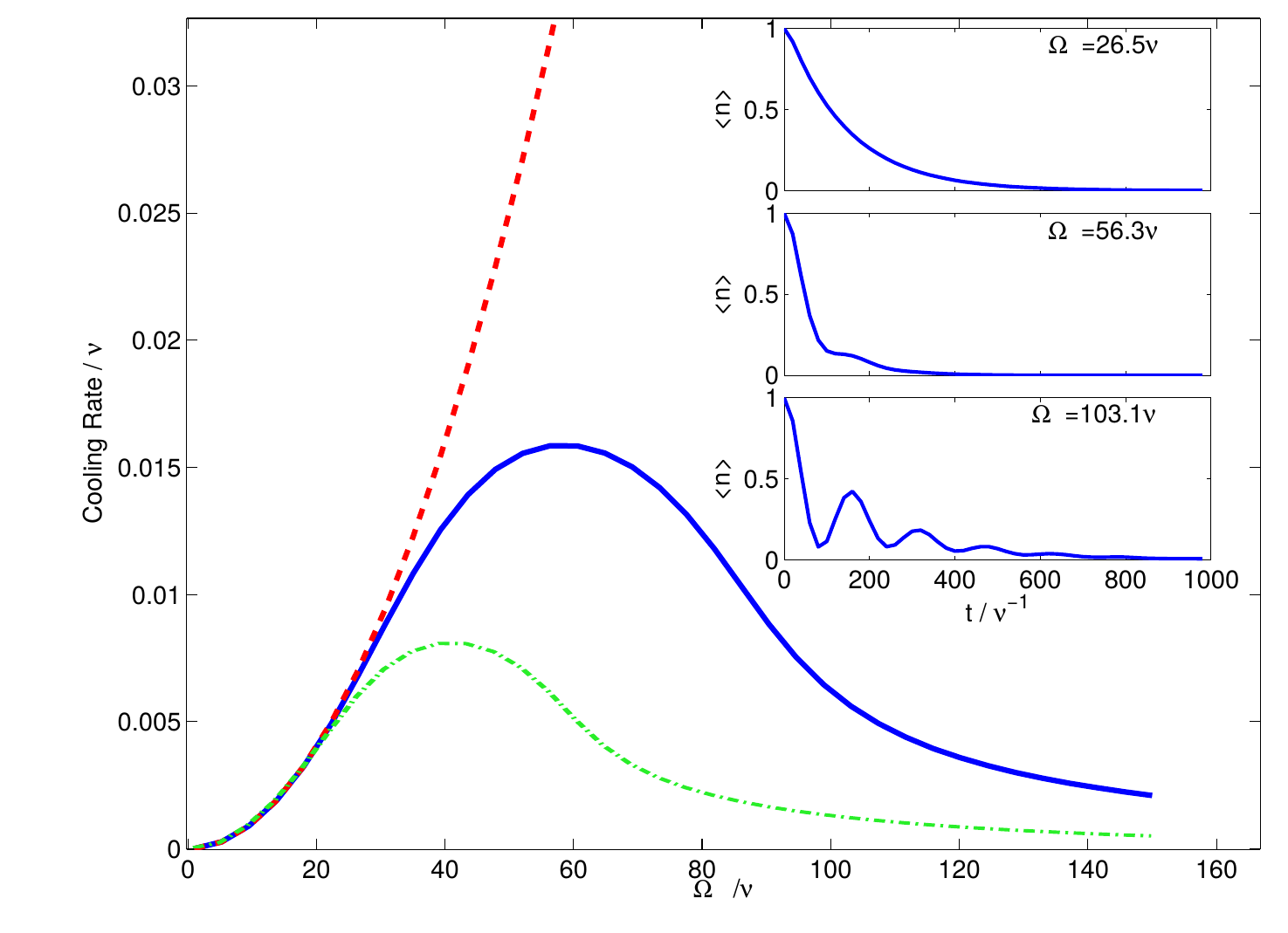}
\end{center}
\caption[Analytical versus numerical predictions for the cooling rate.]{Comparison of the cooling rate $R$ obtained by means of numerical (solid blue line, dot-dashed green line) or analytical (dashed red line) calculations as a function of the Rabi frequency $\Omega$ ($\Gamma=10\nu$, $\Omega_3=-\nu$, $\eta=0.01$). For large values of $\Omega$ the analytical result fails to predict the numerical values. {\em Inset:} Evolution of the average population $\ave n$ as a function of time for three different values of $\W$ starting at a pure motional state at $n=1$.}
\label{robust13}
\end{figure}

As mentioned above, the maximum reachable cooling rates Eq.(\ref{eq:RateLD1}) and Eq.(\ref{eq:RateLD2}) as predicted within the perturbative aproach are limited in practice. The range of validity of the adiabatic treatment used here is constrained to regimes where the coupling between the motional and electronic degrees of freedom is weaker than the effective dissipation rate of the bright states $ \ket + $ and $ \ket e $. Otherwise, mechanical and electrical degrees of freedom couple coherently and the electronic dynamics cannot be averaged out and disentangled from the vibrational dynamics. This is shown in Fig.(\ref{robust13}) for a particular set of parameters, where the analytical result Eq.(\ref{eq:Robustspectraldensity}) predicts the numerical cooling rate correctly only for small values of $\Omega$. The failure of the perturbative prediction can be associated to the appearance of coherence between the motional and electronic degrees of freedom. In particular, for a motional state initially on the first Fock level, coherent oscillations in the cooling process are observed as $\Omega$ is increased. These oscillations are damped at a slower rate than the perturbative approach predict, and they mark the inflection point of the numerical simulation, from where an increase of $\Omega$ corresponds to an actual decrease of overal cooling rate. 
Numerical investigation of our cooling scheme indicates that the inflection point in the rate generally occurs at larger rates as compared to EIT or Stark shift cooling, as shown in Fig.(\ref{robust13}). This may be due to the cancellation to first order of the blue sideband absorption and further investigation is required for the analysis of this behavior.

\section{Experimental Implementation \label{sec:implementation}}

The experimental implementation of the present scheme is possible with state of the art optical systems.  Under the assumption that the transition between the ground and metastable levels is in the microwave range and the ground to excited state corresponds to an optical transition, direct implementation will produce very disparate Lamb-Dicke parameters $\eta$ and $\eta_3$. This is irrelevant for the case associated to Eq.(\ref{EITCond}), since that implies $\eta_3=0$. Nevertheless, for the case Eq.(\ref{SShCond}) the condition for the cancellation of the blue sideband Eq.(\ref{RobustCond}) imposes $\eta_3=4\eta$. This demands a strategy to enhance the Lamb-Dicke parameter associated to the microwave coupling. This is possible from an effective picture and two examples are presented in this section. The first one involves the use of two laser beams in a highly detuned Raman configuration. The second one is performed with microwave driving and Lamb-Dicke enhancement by means of magnetic gradients.

\subsection{Highly detuned Raman beams}

\begin{figure}
\begin{center}
\includegraphics[width=0.8\columnwidth]{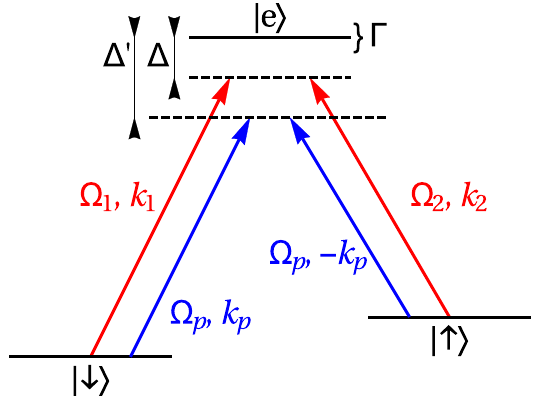}
\end{center}
\caption{Realization of the direct coupling between $\ket{\downarrow}$ and $\ket{uparrow}$ by means of an additional Raman coupling of two beams with Rabi frequency $\W_p$ and opposed projections of their wavevectors $k_p$. The detuning of both couplings with respect to $\ket{e}$ is $\Delta'$.}
\label{fig:implem}
\end{figure}

The direct driving of a microwave transition has an associated Lamb-Dicke parameter which is orders of magnitude smaller than that of an optical frequency. In order to increase its value, we propose to derive coupling 3 from the effect of an additional optical coupling, thus directly ensuring a Lamb-Dicke parameter of similar order to that of couplings 1 and 2. As is shown in Fig.(\ref{fig:implem}) the implementation consists on an additional Raman pair with Rabi frequency $\Omega_p$ and opposing wavevector projections $k_p$. Its detuning $\Delta'$ is chosen to be so large that it effectively decouples from the excited state for all practical purposes. Hence, adiabatic elimination of the upper level is possible, which yields the effective single coupling between  $|\uparrow\rangle$ and $|\downarrow\rangle$. The relationships between our effective parameters $\Omega_3$ and $\eta_3$ and the physical values $\Omega_{p}$ and $\eta_{p}=k_p x_0$ are derived in appendix \ref{sec:effective}, and it is found that, for sufficiently large detunings ($\Delta'\gg\nu$) and in the Lamb-Dicke regime,
\be
\begin{split}
\Omega_3&=\frac{\Omega_p^2}{2\Delta'},\\
\eta_3&=2\eta_{p},
\end{split}
\ee
so that Lamb-Dicke parameters in the optical range may be achieved.

\begin{figure}
\begin{center}
    \includegraphics[width=\columnwidth]{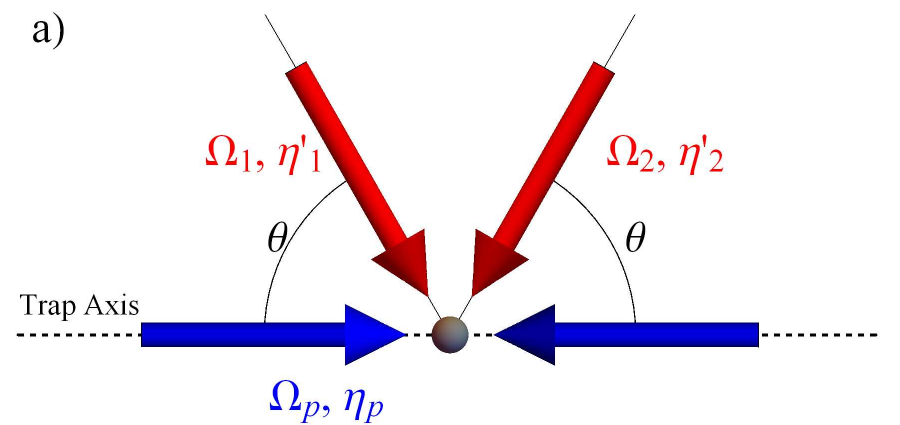}
\end{center}
\begin{center}
    \includegraphics[width=\columnwidth]{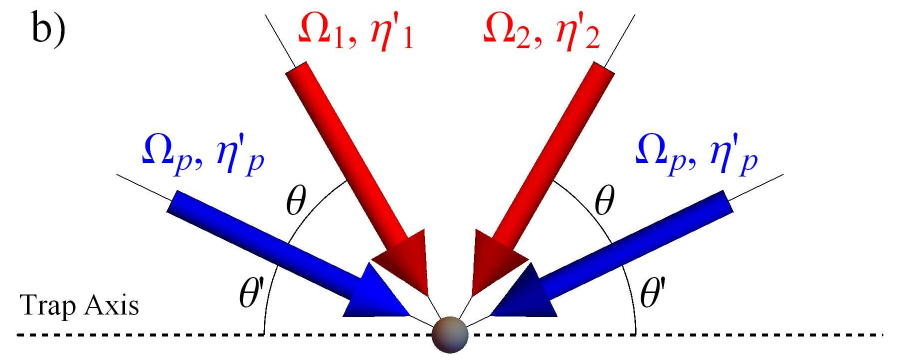}
\end{center}
   \caption[Geometric diagram for cooling in isotropic traps.]{(a) Physical realization of $\Omega_1$, $\W_2$ and $\Omega_3$ couplings, for cooling in the trap axis, as two Raman pair of beams characterized by their respective Rabi frequencies $\Omega_{1,2}$ and $\Omega_p$ and their Lamb-Dicke parameters $\eta'_{1,2}$ and $\eta_p$. (b) Geometrical definition of angle $\theta'$ for an alternative implementation in a trap with reduced optical access.}
\label{fig:polar}
\end{figure}

Condition Eq.(\ref{RobustCond}) may then be fulfilled by taking into account some geometric considerations. By tilting the beams for couplings 1 and 2 an angle $\theta$ with respect to the trap/cooling axis as shown in Fig.(\ref{fig:polar}-a), $\eta_{1,2}=\pm\eta'_{1,2}\cos\theta$, where $\eta'_{1,2}$ are the Lamb-Dicke parameters associated to the axis of propagation of the beams. This provides us with a simple geometrical degree of freedom for the adjustment of the ratio $\eta_3$ to $\eta$.
Even though $\Delta\neq\Delta'$, we use the approximation $\eta'_{1}\simeq\eta'_{2}\simeq\eta_p$ for the sake of simplicity, which is justified in the optical regime. Under this assumption $\frac{2\eta}{\eta_3}=\cos\theta$. By relating this to the resonance Rabi frequency of Eq.(\ref{RobustCond}), $ \cos\theta=\frac{\Omega_3}{\nu+\Omega_3}$. Furthermore, under condition Eq.(\ref{SShCond}), this implies $\theta\simeq 60^\mathrm{o}$.

For the case of ion traps with constrained optical access, other configurations may be considered. For instance, the $p$ Raman pair may be additionally tilted an angle $\theta'$ with respect to the cooling axis, as shown in Fig.(\ref{fig:polar}-b). The ratio of Lamb-Dicke parameters on this axis is $\dfrac{\eta_3}{2\eta}=\dfrac{\cos\theta'}{\cos(\theta+\theta')}$. The optimal cooling condition Eq.(\ref{RobustCond}) is fulfilled for the following angle
\be
\theta'=\arctan\dfrac{(\Omega_3+\nu)\cos\theta-\Omega_3}{(\Omega_3+\nu)\sin\theta},
\label{optimax}
\ee
which is always well defined for any given $\W_3$ and $\frac{\pi}{2}>\theta>0$.

\subsection{Magnetic gradients}
Alternatively, a magnetic gradient can be applied, as is proposed and implemented in a number of recent works in the context of quantum gate implementation \cite{Mintert2001,Timoney2011,Mikelsons2015,Weidt2016,Arrazola2018} and laser cooling \cite{Albrecht2011}. In this system the magnetic gradients create a position dependent Zeeman shift that can be expressed as
\be
\hbar\zeta \left(\ket{\uparrow}\bra{\uparrow} - \ket{\downarrow}\bra{\downarrow} \right) \left(b+b^\dagger\right),
\ee
where $\zeta$ is proportional to the difference in the energy gradient of each level, i.e. $\zeta=x_0|\partial_x \w_\uparrow-\partial_x \w_\downarrow|$, with $x$ the direction of the trap axis. Additionally, the two level system is driven using a microwave of the form $\Omega_3 \left(\ket{\downarrow}\bra{\uparrow}+ H.c.\right) \cos \left(\omega_3 t-\phi_3\right)$, which introduces a crucial degree of control with respect to \cite{Albrecht2011}. After the relevant interaction picture transformation and rotating wave approximation, a rotation to the $\left\{\ket{+},\ket{-}\right\}$ basis yields a very similar Hamiltonian as Eq.(\ref{Ham}), with a modification only of the electronic operator of the first order Hamiltonian, which becomes
\be
\begin{split}
\sigma_\eta =\frac{1}{2}\Big(&i\bar\eta\Omega \ket{e} \bra{+}+ i\eta\Omega e^{i\phi_1} \ket{e} \bra{-} \\
& +2\zeta e^{i\phi_1} \ket{+} \bra{-} + H.c. \Big).
\end{split}\ee
This allows us to recover most of the above results with the substitution $\eta_3=\frac{2\zeta}{\W_3}$.

The range of practically achievable values of the effective Lamb-Dicke parameter is comparable to that of optical transitions \cite{Mintert2001} and is mainly limited by the intensity of the magnetic field gradient $|\nabla B|$. For small fields and electronic levels of small magnetic moment, the assumption is in order that $\zeta$ is directly proportional to the Bohr magneton $\mu_B$ and the field gradient
\be
\zeta\simeq x_0\frac{\mu_B |\nabla B|}{\hbar}.
\ee
For the operative conditions expressed by the combination of Eq.(\ref{RobustCond}) and Eq.(\ref{SShCond}), $\zeta=2\nu\eta$. This provides an aproximation to the necessary magnetic gradients $|\nabla B|\simeq\frac{4\hbar\nu k_1}{\mu_B}$. For an optical transition of about $500\unit{nm}$ and a trap of a few MHz, an estimate gradient of about $5\frac{\unit T}{\unit cm}$ is obtained. Although challenging, this is possible as has been demonstrated in \cite{Reichel1999}.

\section{Multimode Cooling \label{sec:multiaxial}}

\begin{figure}
\begin{center}
\includegraphics[width=\columnwidth]{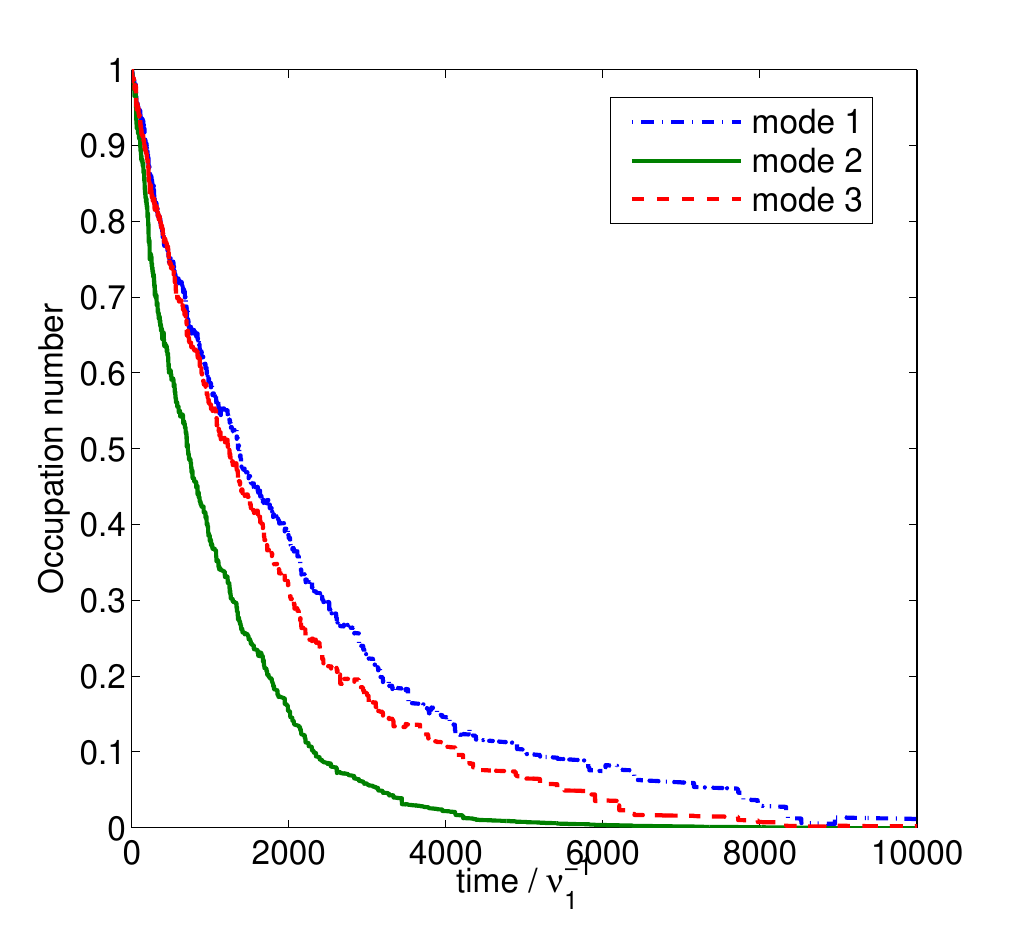}
\end{center}
\caption[Three mode cooling where mode 2 is addressed.]{Occupation number of the three longitudinal normal modes of a chain of three identical ions under double dark state cooling. The laser parameters are set to address the second mode of frequency $\nu_2$. While the cooling is fast and reaches a low temperature in all modes, neighbouring modes 1 and 3 experience a slight decrease in cooling efficiency ($\Omega_3=\nu_2$, $\Omega=3\nu_2$, $\Delta=\W_3/2$, $\Gamma=10\nu_2$, $\eta=0.01$, with $\nu_2$ the eigenfrequency of the second mode and trap strength chosen such that $\nu_1=\nu$).}\label{fig:multim}
\end{figure}

So far, one-dimensional cooling has been discussed. Nevertheless, the flexibility of the experimental implementations discussed above facilitates simultaneous cooling of several modes. The robustness of the scheme makes it possible to cool even if the parameters do not fully satisfy the optimal conditions. This can be utilized both in the context of 3D cooling of a single ion and in the cooling of an ion chain.

\begin{figure}[t]
\begin{center}
\includegraphics[width=\columnwidth]{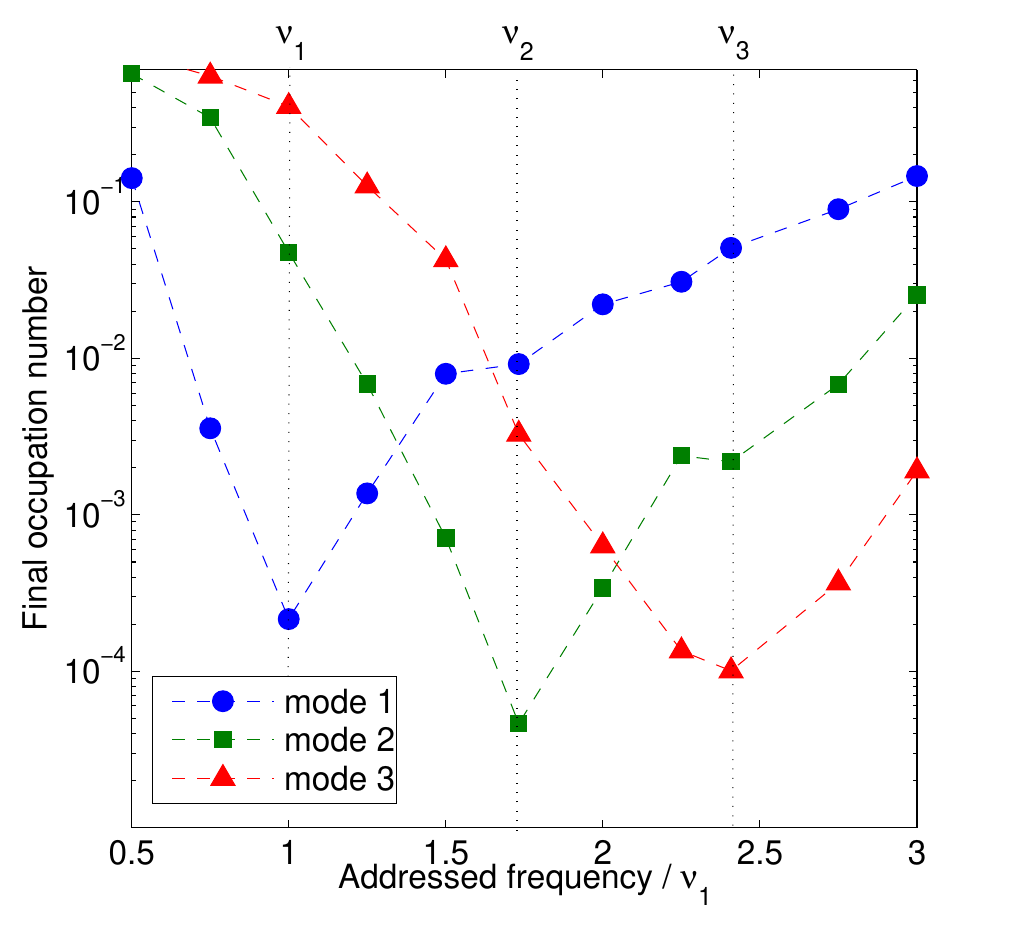}
\end{center}
\caption[Three mode cooling final occupation number.]{Final occupation number of the three longitudinal normal modes as a function of the addressed frequency $\nu_0$. Optimal values for each mode are obtained when the addressed frequency coincides with the eigenfrequency of the corresponding mode. ($\Omega_3=\nu_0$, $\Omega=3\nu_0$, $\Delta=\W_3/2$, $\Gamma=10\nu_2$, $\eta=0.01$ and trap strength chosen such that $\nu_1=\nu$).}\label{fig:chainsweep}
\end{figure}
\begin{figure}
\begin{center}
\includegraphics[width=\columnwidth]{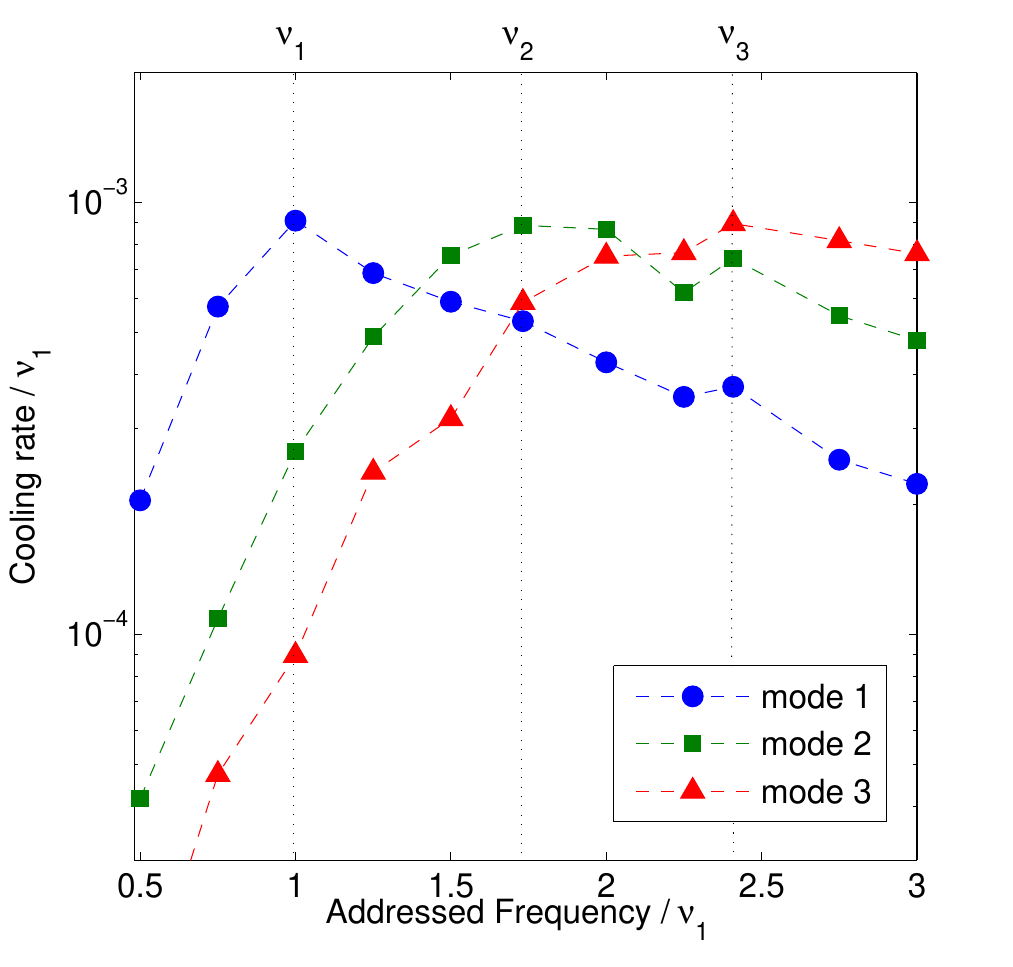}
\end{center}
\caption[Three mode cooling rate.]{Cooling rate of the three longitudinal normal modes as a function of the addressed frequency $\nu_0$. Optimal values for each mode are obtained when the addressed frequency coincides with the eigenfrequency of the corresponding mode. ($\Omega_3=\nu_0$, $\Omega=3\nu_0$, $\Delta=\W_3/2$, $\Gamma=10\nu_2$, $\eta=0.01$ and trap strength chosen such that $\nu_1=\nu$).}\label{fig:chainsweeprate}
\end{figure}

\subsection{Cooling of ion chains \label{sec:multimode}}

The motion of a one dimensional chain of $N$ ions can be described by $N$ normal modes, each characterized by an eigenfrequency $\nu_i$. Cooling for all modes may be achieved by adjusting the parameters such that both peaks of the absorption spectrum are located on the region spanned by the $N$ eigenfrequencies, for instance by the choice $\Delta=\W_3$, $\W_3=\nu_0$ and $\W<\Gamma$, where $\nu_0$ is the most central of the eigenfrequencies $\nu_i$ of the chain. Similarly, condition Eq.(\ref{RobustCond}) is adjusted with reference to the same $\nu_0$.  This is the case presented by the simulation results shown in Fig.(\ref{fig:multim}), where the parameters are set to address mode 2. All modes show roughly the same cooling rate and final temperature, although mode 2 achieves the best performance due to the total cancellation of the blue sideband. A Monte Carlo wave-function simulation method \cite{Molmer1993,Plenio1998} has been used.

When the addressed frequency does not coincide with the central mode, the rate and the final occupation numbers are affected. With the intention to assess this effect, several additional simulations have been run with parameters adjusted to different values of the frequency $\nu_0$. In Fig.(\ref{fig:chainsweep}) and Fig.(\ref{fig:chainsweeprate}) the final occupation number and the cooling rate results of independent simulations are displayed as a function of the addressed frequency $\nu_0$ of each simulation. It shows that optimal values for a particular mode $i$ are mostly obtained when the addressed frequency $\nu_0$ coincides with its frequency $\nu_i$. Nevertheless, the robustness of the scheme guarantees good performance even for distant modes.

As discussed above, with EIT cooling it is not possible to place both peaks of the absorption spectrum on frequencies larger than that of the dark state. Therefore, the neighbouring modes would experience a less efficient cooling with respect to the addressed mode than it is possible with the current proposal.

\subsection{3D cooling}
Finally, it is possible to address the transversal modes in addition to the longitudinal ones. In particular, 3D cooling can be achieved by rearranging the beams for couplings 1 and 2 such that they project onto the additional axes.
 In such situation, double EIT cooling can only be achieved in one axis, but efficient cooling can be achieved for the other two following the strategy presented above for ion chains.


\section{Conclusions}

A detailed study of an efficient and fast laser cooling scheme for trapped three level systems has been presented. It has been shown that a particular setting of the Lamb-Dicke parameters ratio combines two underlying cooling schemes so that the blue sideband can be effectively canceled. Thus, total occupation of the ground state is reached up to second order in the Lamb-Dicke expansion. Being both a fast and a robust cooling scheme, two different experimental implementations have been proposed that equivalently perform the scheme. Specific geometric configurations of laser beams would be able to also cool the two remaining axes, thus converting it into a 3D cooling scheme. Its suitability for cooling several ions in a trap has been explored.

\acknowledgments
This work was supported by the AXA Research Fund, the ERC consolidator grant QRES, the ERC Synergy grant BioQ and the Alexander von Humboldt foundation.

\bibliography{RobustLP}

\begin{thebibliography}{29}%
\makeatletter
\providecommand \@ifxundefined [1]{%
 \@ifx{#1\undefined}
}%
\providecommand \@ifnum [1]{%
 \ifnum #1\expandafter \@firstoftwo
 \else \expandafter \@secondoftwo
 \fi
}%
\providecommand \@ifx [1]{%
 \ifx #1\expandafter \@firstoftwo
 \else \expandafter \@secondoftwo
 \fi
}%
\providecommand \natexlab [1]{#1}%
\providecommand \enquote  [1]{``#1''}%
\providecommand \bibnamefont  [1]{#1}%
\providecommand \bibfnamefont [1]{#1}%
\providecommand \citenamefont [1]{#1}%
\providecommand \href@noop [0]{\@secondoftwo}%
\providecommand \href [0]{\begingroup \@sanitize@url \@href}%
\providecommand \@href[1]{\@@startlink{#1}\@@href}%
\providecommand \@@href[1]{\endgroup#1\@@endlink}%
\providecommand \@sanitize@url [0]{\catcode `\\12\catcode `\$12\catcode
  `\&12\catcode `\#12\catcode `\^12\catcode `\_12\catcode `\%12\relax}%
\providecommand \@@startlink[1]{}%
\providecommand \@@endlink[0]{}%
\providecommand \url  [0]{\begingroup\@sanitize@url \@url }%
\providecommand \@url [1]{\endgroup\@href {#1}{\urlprefix }}%
\providecommand \urlprefix  [0]{URL }%
\providecommand \Eprint [0]{\href }%
\providecommand \doibase [0]{http://dx.doi.org/}%
\providecommand \selectlanguage [0]{\@gobble}%
\providecommand \bibinfo  [0]{\@secondoftwo}%
\providecommand \bibfield  [0]{\@secondoftwo}%
\providecommand \translation [1]{[#1]}%
\providecommand \BibitemOpen [0]{}%
\providecommand \bibitemStop [0]{}%
\providecommand \bibitemNoStop [0]{.\EOS\space}%
\providecommand \EOS [0]{\spacefactor3000\relax}%
\providecommand \BibitemShut  [1]{\csname bibitem#1\endcsname}%
\let\auto@bib@innerbib\@empty
\bibitem [{\citenamefont {Chu}(1998)}]{Chu1998}%
  \BibitemOpen
  \bibfield  {author} {\bibinfo {author} {\bibfnamefont {S.}~\bibnamefont
  {Chu}},\ }\href
  {http://www.nobelprize.org/nobel{\_}prizes/physics/laureates/1997/chu-lecture.pdf}
  {\bibfield  {journal} {\bibinfo  {journal} {Rev. Mod. Phys.}\ }\textbf
  {\bibinfo {volume} {70}},\ \bibinfo {pages} {685} (\bibinfo {year}
  {1998})}\BibitemShut {NoStop}%
\bibitem [{\citenamefont {Cohen-Tannoudji}(1998)}]{Cohen-Tannoudji1998}%
  \BibitemOpen
  \bibfield  {author} {\bibinfo {author} {\bibfnamefont {C.}~\bibnamefont
  {Cohen-Tannoudji}},\ }\href {\doibase 10.1238/Physica.Topical.076a00033}
  {\bibfield  {journal} {\bibinfo  {journal} {Rev. Mod. Phys.}\ }\textbf
  {\bibinfo {volume} {70}},\ \bibinfo {pages} {707} (\bibinfo {year}
  {1998})}\BibitemShut {NoStop}%
\bibitem [{\citenamefont {Phillips}(1998)}]{Phillips1998}%
  \BibitemOpen
  \bibfield  {author} {\bibinfo {author} {\bibfnamefont {W.~D.}\ \bibnamefont
  {Phillips}},\ }\href {\doibase 10.1103/RevModPhys.70.721} {\bibfield
  {journal} {\bibinfo  {journal} {Rev. Mod. Phys.}\ }\textbf {\bibinfo {volume}
  {70}},\ \bibinfo {pages} {721} (\bibinfo {year} {1998})}\BibitemShut
  {NoStop}%
\bibitem [{\citenamefont {H{\"{a}}nsch}\ and\ \citenamefont
  {Schawlow}(1975)}]{Hansch1975}%
  \BibitemOpen
  \bibfield  {author} {\bibinfo {author} {\bibfnamefont {T.~W.}\ \bibnamefont
  {H{\"{a}}nsch}}\ and\ \bibinfo {author} {\bibfnamefont {A.~L.}\ \bibnamefont
  {Schawlow}},\ }\href {\doibase 10.1016/0030-4018(75)90159-5} {\bibfield
  {journal} {\bibinfo  {journal} {Opt. Commun.}\ }\textbf {\bibinfo {volume}
  {13}},\ \bibinfo {pages} {68} (\bibinfo {year} {1975})}\BibitemShut {NoStop}%
\bibitem [{\citenamefont {Wineland}\ and\ \citenamefont
  {Dehmelt}(1975)}]{Wineland1975}%
  \BibitemOpen
  \bibfield  {author} {\bibinfo {author} {\bibfnamefont {D.~J.}\ \bibnamefont
  {Wineland}}\ and\ \bibinfo {author} {\bibfnamefont {H.}~\bibnamefont
  {Dehmelt}},\ }\href@noop {} {\bibfield  {journal} {\bibinfo  {journal} {Bull.
  Am. Phys. Soc.}\ }\textbf {\bibinfo {volume} {20}},\ \bibinfo {pages} {637}
  (\bibinfo {year} {1975})}\BibitemShut {NoStop}%
\bibitem [{\citenamefont {Wineland}\ \emph {et~al.}(1978)\citenamefont
  {Wineland}, \citenamefont {Drullinger},\ and\ \citenamefont
  {Walls}}]{Wineland1978}%
  \BibitemOpen
  \bibfield  {author} {\bibinfo {author} {\bibfnamefont {D.~J.}\ \bibnamefont
  {Wineland}}, \bibinfo {author} {\bibfnamefont {R.~E.}\ \bibnamefont
  {Drullinger}}, \ and\ \bibinfo {author} {\bibfnamefont {F.~L.}\ \bibnamefont
  {Walls}},\ }\href {\doibase 10.1103/PhysRevLett.40.1639} {\bibfield
  {journal} {\bibinfo  {journal} {Phys. Rev. Lett.}\ }\textbf {\bibinfo
  {volume} {40}},\ \bibinfo {pages} {1639} (\bibinfo {year}
  {1978})}\BibitemShut {NoStop}%
\bibitem [{\citenamefont {Aspect}\ \emph {et~al.}(1988)\citenamefont {Aspect},
  \citenamefont {Arimondo}, \citenamefont {Kaiser}, \citenamefont
  {Vansteenkiste},\ and\ \citenamefont {Cohen-Tannoudji}}]{Aspect1988}%
  \BibitemOpen
  \bibfield  {author} {\bibinfo {author} {\bibfnamefont {A.}~\bibnamefont
  {Aspect}}, \bibinfo {author} {\bibfnamefont {E.}~\bibnamefont {Arimondo}},
  \bibinfo {author} {\bibfnamefont {R.}~\bibnamefont {Kaiser}}, \bibinfo
  {author} {\bibfnamefont {N.}~\bibnamefont {Vansteenkiste}}, \ and\ \bibinfo
  {author} {\bibfnamefont {C.}~\bibnamefont {Cohen-Tannoudji}},\ }\href@noop {}
  {\bibfield  {journal} {\bibinfo  {journal} {Phys. Rev. Lett.}\ }\textbf
  {\bibinfo {volume} {61}},\ \bibinfo {pages} {826} (\bibinfo {year}
  {1988})}\BibitemShut {NoStop}%
\bibitem [{\citenamefont {Dum}\ \emph {et~al.}(1994)\citenamefont {Dum},
  \citenamefont {Marte}, \citenamefont {Pellizzari},\ and\ \citenamefont
  {Zoller}}]{Dum1994}%
  \BibitemOpen
  \bibfield  {author} {\bibinfo {author} {\bibfnamefont {R.}~\bibnamefont
  {Dum}}, \bibinfo {author} {\bibfnamefont {P.}~\bibnamefont {Marte}}, \bibinfo
  {author} {\bibfnamefont {T.}~\bibnamefont {Pellizzari}}, \ and\ \bibinfo
  {author} {\bibfnamefont {P.}~\bibnamefont {Zoller}},\ }\href {\doibase
  10.1103/PhysRevLett.73.2829} {\bibfield  {journal} {\bibinfo  {journal}
  {Phys. Rev. Lett.}\ }\textbf {\bibinfo {volume} {73}},\ \bibinfo {pages}
  {2829} (\bibinfo {year} {1994})}\BibitemShut {NoStop}%
\bibitem [{\citenamefont {Marzoli}\ \emph {et~al.}(1994)\citenamefont
  {Marzoli}, \citenamefont {Cirac}, \citenamefont {Blatt},\ and\ \citenamefont
  {Zoller}}]{Marzoli1994}%
  \BibitemOpen
  \bibfield  {author} {\bibinfo {author} {\bibfnamefont {I.}~\bibnamefont
  {Marzoli}}, \bibinfo {author} {\bibfnamefont {J.~I.}\ \bibnamefont {Cirac}},
  \bibinfo {author} {\bibfnamefont {R.}~\bibnamefont {Blatt}}, \ and\ \bibinfo
  {author} {\bibfnamefont {P.}~\bibnamefont {Zoller}},\ }\href {\doibase
  10.1063/1.881076} {\bibfield  {journal} {\bibinfo  {journal} {Phys. Rev. A}\
  }\textbf {\bibinfo {volume} {49}},\ \bibinfo {pages} {2771} (\bibinfo {year}
  {1994})}\BibitemShut {NoStop}%
\bibitem [{\citenamefont {Monroe}\ \emph {et~al.}(1995)\citenamefont {Monroe},
  \citenamefont {Meekhof}, \citenamefont {King}, \citenamefont {Jefferts},
  \citenamefont {Itano}, \citenamefont {Wineland},\ and\ \citenamefont
  {Gould}}]{Monroe1995}%
  \BibitemOpen
  \bibfield  {author} {\bibinfo {author} {\bibfnamefont {C.}~\bibnamefont
  {Monroe}}, \bibinfo {author} {\bibfnamefont {D.~M.}\ \bibnamefont {Meekhof}},
  \bibinfo {author} {\bibfnamefont {B.~E.}\ \bibnamefont {King}}, \bibinfo
  {author} {\bibfnamefont {S.~R.}\ \bibnamefont {Jefferts}}, \bibinfo {author}
  {\bibfnamefont {W.~M.}\ \bibnamefont {Itano}}, \bibinfo {author}
  {\bibfnamefont {D.~J.}\ \bibnamefont {Wineland}}, \ and\ \bibinfo {author}
  {\bibfnamefont {P.}~\bibnamefont {Gould}},\ }\href {\doibase
  10.1103/PhysRevLett.75.4011} {\bibfield  {journal} {\bibinfo  {journal}
  {Phys. Rev. Lett.}\ }\textbf {\bibinfo {volume} {75}},\ \bibinfo {pages}
  {4011} (\bibinfo {year} {1995})}\BibitemShut {NoStop}%
\bibitem [{\citenamefont {Fleischhauer}\ \emph {et~al.}(2005)\citenamefont
  {Fleischhauer}, \citenamefont {Imamoglu},\ and\ \citenamefont
  {Marangos}}]{Fleischhauer2005}%
  \BibitemOpen
  \bibfield  {author} {\bibinfo {author} {\bibfnamefont {M.}~\bibnamefont
  {Fleischhauer}}, \bibinfo {author} {\bibfnamefont {A.}~\bibnamefont
  {Imamoglu}}, \ and\ \bibinfo {author} {\bibfnamefont {J.~P.}\ \bibnamefont
  {Marangos}},\ }\href {\doibase 10.1103/RevModPhys.77.633} {\bibfield
  {journal} {\bibinfo  {journal} {Rev. Mod. Phys.}\ }\textbf {\bibinfo {volume}
  {77}},\ \bibinfo {pages} {633} (\bibinfo {year} {2005})}\BibitemShut
  {NoStop}%
\bibitem [{\citenamefont {Morigi}\ \emph {et~al.}(2000)\citenamefont {Morigi},
  \citenamefont {Eschner},\ and\ \citenamefont {Keitel}}]{Morigi2000}%
  \BibitemOpen
  \bibfield  {author} {\bibinfo {author} {\bibfnamefont {G.}~\bibnamefont
  {Morigi}}, \bibinfo {author} {\bibfnamefont {J.}~\bibnamefont {Eschner}}, \
  and\ \bibinfo {author} {\bibfnamefont {C.}~\bibnamefont {Keitel}},\ }\href
  {\doibase 10.1103/PhysRevLett.85.4458} {\bibfield  {journal} {\bibinfo
  {journal} {Phys. Rev. Lett.}\ }\textbf {\bibinfo {volume} {85}},\ \bibinfo
  {pages} {4458} (\bibinfo {year} {2000})}\BibitemShut {NoStop}%
\bibitem [{\citenamefont {Roos}\ \emph {et~al.}(2000)\citenamefont {Roos},
  \citenamefont {Leibfried}, \citenamefont {Mundt}, \citenamefont
  {Schmidt-Kaler}, \citenamefont {Eschner},\ and\ \citenamefont
  {Blatt}}]{Roos2000}%
  \BibitemOpen
  \bibfield  {author} {\bibinfo {author} {\bibfnamefont {C.~F.}\ \bibnamefont
  {Roos}}, \bibinfo {author} {\bibfnamefont {D.}~\bibnamefont {Leibfried}},
  \bibinfo {author} {\bibfnamefont {A.}~\bibnamefont {Mundt}}, \bibinfo
  {author} {\bibfnamefont {F.}~\bibnamefont {Schmidt-Kaler}}, \bibinfo {author}
  {\bibfnamefont {J.}~\bibnamefont {Eschner}}, \ and\ \bibinfo {author}
  {\bibfnamefont {R.}~\bibnamefont {Blatt}},\ }\href {\doibase
  10.1103/PhysRevLett.85.5547} {\bibfield  {journal} {\bibinfo  {journal}
  {Phys. Rev. Lett.}\ }\textbf {\bibinfo {volume} {85}},\ \bibinfo {pages}
  {5547} (\bibinfo {year} {2000})}\BibitemShut {NoStop}%
\bibitem [{\citenamefont {Morigi}(2003)}]{Morigi2003}%
  \BibitemOpen
  \bibfield  {author} {\bibinfo {author} {\bibfnamefont {G.}~\bibnamefont
  {Morigi}},\ }\href {\doibase 10.1103/PhysRevA.67.033402} {\bibfield
  {journal} {\bibinfo  {journal} {Phys. Rev. A}\ }\textbf {\bibinfo {volume}
  {67}},\ \bibinfo {pages} {033402} (\bibinfo {year} {2003})},\ \Eprint
  {http://arxiv.org/abs/0211043} {0211043 [quant-ph]} \BibitemShut {NoStop}%
\bibitem [{\citenamefont {Retzker}\ and\ \citenamefont
  {Plenio}(2007)}]{Retzker2007}%
  \BibitemOpen
  \bibfield  {author} {\bibinfo {author} {\bibfnamefont {A.}~\bibnamefont
  {Retzker}}\ and\ \bibinfo {author} {\bibfnamefont {M.~B.}\ \bibnamefont
  {Plenio}},\ }\href {\doibase 10.1088/1367-2630/9/8/279} {\bibfield  {journal}
  {\bibinfo  {journal} {New J. Phys.}\ }\textbf {\bibinfo {volume} {9}},\
  \bibinfo {pages} {279} (\bibinfo {year} {2007})}\BibitemShut {NoStop}%
\bibitem [{\citenamefont {Jonathan}\ \emph {et~al.}(2000)\citenamefont
  {Jonathan}, \citenamefont {Plenio},\ and\ \citenamefont
  {Knight}}]{Jonathan2000}%
  \BibitemOpen
  \bibfield  {author} {\bibinfo {author} {\bibfnamefont {D.}~\bibnamefont
  {Jonathan}}, \bibinfo {author} {\bibfnamefont {M.~B.}\ \bibnamefont
  {Plenio}}, \ and\ \bibinfo {author} {\bibfnamefont {P.~L.}\ \bibnamefont
  {Knight}},\ }\href {\doibase 10.1103/PhysRevA.62.042307} {\bibfield
  {journal} {\bibinfo  {journal} {Phys. Rev. A}\ }\textbf {\bibinfo {volume}
  {62}},\ \bibinfo {pages} {10} (\bibinfo {year} {2000})}\BibitemShut {NoStop}%
\bibitem [{\citenamefont {Evers}\ and\ \citenamefont
  {Keitel}(2004)}]{Evers2004}%
  \BibitemOpen
  \bibfield  {author} {\bibinfo {author} {\bibfnamefont {J.}~\bibnamefont
  {Evers}}\ and\ \bibinfo {author} {\bibfnamefont {C.~H.}\ \bibnamefont
  {Keitel}},\ }\href {\doibase 10.1209/epl/i2004-10207-5} {\bibfield  {journal}
  {\bibinfo  {journal} {Europhys. Lett.}\ }\textbf {\bibinfo {volume} {68}},\
  \bibinfo {pages} {370} (\bibinfo {year} {2004})}\BibitemShut {NoStop}%
\bibitem [{\citenamefont {Cerrillo}\ \emph {et~al.}(2010)\citenamefont
  {Cerrillo}, \citenamefont {Retzker},\ and\ \citenamefont
  {Plenio}}]{Cerrillo2010}%
  \BibitemOpen
  \bibfield  {author} {\bibinfo {author} {\bibfnamefont {J.}~\bibnamefont
  {Cerrillo}}, \bibinfo {author} {\bibfnamefont {A.}~\bibnamefont {Retzker}}, \
  and\ \bibinfo {author} {\bibfnamefont {M.~B.}\ \bibnamefont {Plenio}},\
  }\href {\doibase 10.1103/PhysRevLett.104.043003} {\bibfield  {journal}
  {\bibinfo  {journal} {Phys. Rev. Lett.}\ }\textbf {\bibinfo {volume} {104}},\
  \bibinfo {pages} {043003} (\bibinfo {year} {2010})}\BibitemShut {NoStop}%
\bibitem [{\citenamefont {Scharnhorst}\ \emph {et~al.}(2017)\citenamefont
  {Scharnhorst}, \citenamefont {Cerrillo}, \citenamefont {Kramer},
  \citenamefont {Leroux}, \citenamefont {W{\"{u}}bbena}, \citenamefont
  {Retzker},\ and\ \citenamefont {Schmidt}}]{Scharnhorst2017}%
  \BibitemOpen
  \bibfield  {author} {\bibinfo {author} {\bibfnamefont {N.}~\bibnamefont
  {Scharnhorst}}, \bibinfo {author} {\bibfnamefont {J.}~\bibnamefont
  {Cerrillo}}, \bibinfo {author} {\bibfnamefont {J.}~\bibnamefont {Kramer}},
  \bibinfo {author} {\bibfnamefont {I.~D.}\ \bibnamefont {Leroux}}, \bibinfo
  {author} {\bibfnamefont {J.~B.}\ \bibnamefont {W{\"{u}}bbena}}, \bibinfo
  {author} {\bibfnamefont {A.}~\bibnamefont {Retzker}}, \ and\ \bibinfo
  {author} {\bibfnamefont {P.~O.}\ \bibnamefont {Schmidt}},\ }\href
  {http://arxiv.org/abs/1711.00732} {\  (\bibinfo {year} {2017})},\ \Eprint
  {http://arxiv.org/abs/1711.00732} {arXiv:1711.00732} \BibitemShut {NoStop}%
\bibitem [{\citenamefont {Mintert}\ and\ \citenamefont
  {Wunderlich}(2001)}]{Mintert2001}%
  \BibitemOpen
  \bibfield  {author} {\bibinfo {author} {\bibfnamefont {F.}~\bibnamefont
  {Mintert}}\ and\ \bibinfo {author} {\bibfnamefont {C.}~\bibnamefont
  {Wunderlich}},\ }\href {\doibase 10.1103/PhysRevLett.87.257904} {\bibfield
  {journal} {\bibinfo  {journal} {Phys. Rev. Lett.}\ }\textbf {\bibinfo
  {volume} {87}},\ \bibinfo {pages} {257904} (\bibinfo {year}
  {2001})}\BibitemShut {NoStop}%
\bibitem [{\citenamefont {Timoney}\ \emph {et~al.}(2011)\citenamefont
  {Timoney}, \citenamefont {Baumgart}, \citenamefont {Johanning}, \citenamefont
  {Var{\'{o}}n}, \citenamefont {Plenio}, \citenamefont {Retzker},\ and\
  \citenamefont {Wunderlich}}]{Timoney2011}%
  \BibitemOpen
  \bibfield  {author} {\bibinfo {author} {\bibfnamefont {N.}~\bibnamefont
  {Timoney}}, \bibinfo {author} {\bibfnamefont {I.}~\bibnamefont {Baumgart}},
  \bibinfo {author} {\bibfnamefont {M.}~\bibnamefont {Johanning}}, \bibinfo
  {author} {\bibfnamefont {A.~F.}\ \bibnamefont {Var{\'{o}}n}}, \bibinfo
  {author} {\bibfnamefont {M.~B.}\ \bibnamefont {Plenio}}, \bibinfo {author}
  {\bibfnamefont {A.}~\bibnamefont {Retzker}}, \ and\ \bibinfo {author}
  {\bibfnamefont {C.}~\bibnamefont {Wunderlich}},\ }\href
  {http://dx.doi.org/10.1038/nature10319 http://10.0.4.14/nature10319
  https://www.nature.com/articles/nature10319{\#}supplementary-information}
  {\bibfield  {journal} {\bibinfo  {journal} {Nature}\ }\textbf {\bibinfo
  {volume} {476}},\ \bibinfo {pages} {185} (\bibinfo {year}
  {2011})}\BibitemShut {NoStop}%
\bibitem [{\citenamefont {Mikelsons}\ \emph {et~al.}(2015)\citenamefont
  {Mikelsons}, \citenamefont {Cohen}, \citenamefont {Retzker},\ and\
  \citenamefont {Plenio}}]{Mikelsons2015}%
  \BibitemOpen
  \bibfield  {author} {\bibinfo {author} {\bibfnamefont {G.}~\bibnamefont
  {Mikelsons}}, \bibinfo {author} {\bibfnamefont {I.}~\bibnamefont {Cohen}},
  \bibinfo {author} {\bibfnamefont {A.}~\bibnamefont {Retzker}}, \ and\
  \bibinfo {author} {\bibfnamefont {M.~B.}\ \bibnamefont {Plenio}},\ }\href
  {\doibase 10.1088/1367-2630/17/5/053032} {\bibfield  {journal} {\bibinfo
  {journal} {New J. Phys.}\ }\textbf {\bibinfo {volume} {17}},\ \bibinfo
  {pages} {053032} (\bibinfo {year} {2015})}\BibitemShut {NoStop}%
\bibitem [{\citenamefont {Weidt}\ \emph {et~al.}(2016)\citenamefont {Weidt},
  \citenamefont {Randall}, \citenamefont {Webster}, \citenamefont {Lake},
  \citenamefont {Webb}, \citenamefont {Cohen}, \citenamefont {Navickas},
  \citenamefont {Lekitsch}, \citenamefont {Retzker},\ and\ \citenamefont
  {Hensinger}}]{Weidt2016}%
  \BibitemOpen
  \bibfield  {author} {\bibinfo {author} {\bibfnamefont {S.}~\bibnamefont
  {Weidt}}, \bibinfo {author} {\bibfnamefont {J.}~\bibnamefont {Randall}},
  \bibinfo {author} {\bibfnamefont {S.~C.}\ \bibnamefont {Webster}}, \bibinfo
  {author} {\bibfnamefont {K.}~\bibnamefont {Lake}}, \bibinfo {author}
  {\bibfnamefont {A.~E.}\ \bibnamefont {Webb}}, \bibinfo {author}
  {\bibfnamefont {I.}~\bibnamefont {Cohen}}, \bibinfo {author} {\bibfnamefont
  {T.}~\bibnamefont {Navickas}}, \bibinfo {author} {\bibfnamefont
  {B.}~\bibnamefont {Lekitsch}}, \bibinfo {author} {\bibfnamefont
  {A.}~\bibnamefont {Retzker}}, \ and\ \bibinfo {author} {\bibfnamefont
  {W.~K.}\ \bibnamefont {Hensinger}},\ }\href {\doibase
  10.1103/PhysRevLett.117.220501} {\bibfield  {journal} {\bibinfo  {journal}
  {Phys. Rev. Lett.}\ }\textbf {\bibinfo {volume} {117}},\ \bibinfo {pages}
  {220501} (\bibinfo {year} {2016})}\BibitemShut {NoStop}%
\bibitem [{\citenamefont {Arrazola}\ \emph {et~al.}(2018)\citenamefont
  {Arrazola}, \citenamefont {Casanova}, \citenamefont {Pedernales},
  \citenamefont {Wang}, \citenamefont {Solano},\ and\ \citenamefont
  {Plenio}}]{Arrazola2018}%
  \BibitemOpen
  \bibfield  {author} {\bibinfo {author} {\bibfnamefont {I.}~\bibnamefont
  {Arrazola}}, \bibinfo {author} {\bibfnamefont {J.}~\bibnamefont {Casanova}},
  \bibinfo {author} {\bibfnamefont {J.~S.}\ \bibnamefont {Pedernales}},
  \bibinfo {author} {\bibfnamefont {Z.-Y.}\ \bibnamefont {Wang}}, \bibinfo
  {author} {\bibfnamefont {E.}~\bibnamefont {Solano}}, \ and\ \bibinfo {author}
  {\bibfnamefont {M.~B.}\ \bibnamefont {Plenio}},\ }\href {\doibase
  10.1103/PhysRevA.97.052312} {\bibfield  {journal} {\bibinfo  {journal} {Phys.
  Rev. A}\ }\textbf {\bibinfo {volume} {97}},\ \bibinfo {pages} {052312}
  (\bibinfo {year} {2018})}\BibitemShut {NoStop}%
\bibitem [{\citenamefont {Albrecht}\ \emph {et~al.}(2011)\citenamefont
  {Albrecht}, \citenamefont {Retzker}, \citenamefont {Wunderlich},\ and\
  \citenamefont {Plenio}}]{Albrecht2011}%
  \BibitemOpen
  \bibfield  {author} {\bibinfo {author} {\bibfnamefont {A.}~\bibnamefont
  {Albrecht}}, \bibinfo {author} {\bibfnamefont {A.}~\bibnamefont {Retzker}},
  \bibinfo {author} {\bibfnamefont {C.}~\bibnamefont {Wunderlich}}, \ and\
  \bibinfo {author} {\bibfnamefont {M.~B.}\ \bibnamefont {Plenio}},\ }\href
  {\doibase 10.1088/1367-2630/13/3/033009} {\bibfield  {journal} {\bibinfo
  {journal} {New J. Phys.}\ }\textbf {\bibinfo {volume} {13}},\ \bibinfo
  {pages} {033009} (\bibinfo {year} {2011})}\BibitemShut {NoStop}%
\bibitem [{\citenamefont {Reichel}\ \emph {et~al.}(1999)\citenamefont
  {Reichel}, \citenamefont {H{\"{a}}nsel},\ and\ \citenamefont
  {H{\"{a}}nsch}}]{Reichel1999}%
  \BibitemOpen
  \bibfield  {author} {\bibinfo {author} {\bibfnamefont {J.}~\bibnamefont
  {Reichel}}, \bibinfo {author} {\bibfnamefont {W.}~\bibnamefont
  {H{\"{a}}nsel}}, \ and\ \bibinfo {author} {\bibfnamefont {T.~W.}\
  \bibnamefont {H{\"{a}}nsch}},\ }\href {\doibase 10.1103/PhysRevLett.83.3398}
  {\bibfield  {journal} {\bibinfo  {journal} {Phys. Rev. Lett.}\ }\textbf
  {\bibinfo {volume} {83}},\ \bibinfo {pages} {3398} (\bibinfo {year}
  {1999})}\BibitemShut {NoStop}%
\bibitem [{\citenamefont {M{\o}lmer}\ \emph {et~al.}(1993)\citenamefont
  {M{\o}lmer}, \citenamefont {Castin},\ and\ \citenamefont
  {Dalibard}}]{Molmer1993}%
  \BibitemOpen
  \bibfield  {author} {\bibinfo {author} {\bibfnamefont {K.}~\bibnamefont
  {M{\o}lmer}}, \bibinfo {author} {\bibfnamefont {Y.}~\bibnamefont {Castin}}, \
  and\ \bibinfo {author} {\bibfnamefont {J.}~\bibnamefont {Dalibard}},\ }\href
  {http://www.opticsinfobase.org/abstract.cfm?{\&}id=59382} {\bibfield
  {journal} {\bibinfo  {journal} {J. Opt. Soc. Am. B}\ }\textbf {\bibinfo
  {volume} {10}},\ \bibinfo {pages} {524} (\bibinfo {year} {1993})}\BibitemShut
  {NoStop}%
\bibitem [{\citenamefont {Plenio}\ and\ \citenamefont
  {Knight}(1998)}]{Plenio1998}%
  \BibitemOpen
  \bibfield  {author} {\bibinfo {author} {\bibfnamefont {M.~B.}\ \bibnamefont
  {Plenio}}\ and\ \bibinfo {author} {\bibfnamefont {P.~L.}\ \bibnamefont
  {Knight}},\ }\href {\doibase 10.1103/RevModPhys.70.101} {\bibfield  {journal}
  {\bibinfo  {journal} {Rev. Mod. Phys.}\ }\textbf {\bibinfo {volume} {70}},\
  \bibinfo {pages} {101} (\bibinfo {year} {1998})}\BibitemShut {NoStop}%
\bibitem [{\citenamefont {James}\ and\ \citenamefont
  {Jerke}(2007)}]{James2007}%
  \BibitemOpen
  \bibfield  {author} {\bibinfo {author} {\bibfnamefont {D.~F.}\ \bibnamefont
  {James}}\ and\ \bibinfo {author} {\bibfnamefont {J.}~\bibnamefont {Jerke}},\
  }\href@noop {} {\bibfield  {journal} {\bibinfo  {journal} {Can. J. Phys.}\
  }\textbf {\bibinfo {volume} {85}},\ \bibinfo {pages} {625} (\bibinfo {year}
  {2007})}\BibitemShut {NoStop}%
\end{thebibliography}%

\appendix
\section{Real part of the cooling spectrum \label{sec:spectrumreal} }
The spectrum Eq.(\ref{spectrum}) becomes within the quantum regression theorem
\be
S(\w)= \mathrm{Tr}\left\lbrace\sigma_{\eta}\frac{1}{i\w + \mathcal{L}_0}\sigma_{\eta}\ket{-}\bra{-}\right\rbrace,
\ee
and, due to the property
\be
\mathcal{L}_0
\left(\begin{array}{c}
\ket{e} \bra{-}\\
\ket{+} \bra{-}
\end{array}\right)
=
\mathrm{L}_0
\left(\begin{array}{c}
\ket{e} \bra{-}\\
\ket{+} \bra{-}
\end{array}\right),
\ee
with
\be
\mathrm{L}_0=-i \left(\begin{array}{cc}
\Delta + \frac{\Omega_3}{2} -i \Gamma & \frac{\Omega}{2}\\
\frac{\Omega}{2} & \Omega_3
\end{array}\right),
\ee
can be expressed as the inner product
\be
S(\w)= 
 \bra{\eta} \left( i\w+ \mathrm{L}_0\right) ^{-1}\ket{\eta},
\ee
with $\ket{\eta}=\frac{1}{2}\left(\eta\Omega\ket{e} + \eta_3\Omega_3\ket{+}\right)$, and for $\bra{e}=(1,0)$ and $\bra{+}=(0,1)$. The real part thereof, relevant for the rates $A_\pm$ Eq.(\ref{rates}), may be derived from
\be\begin{split}
2&Re\left[S(\w)\right]=\\
&\bra{\eta} \left[\frac{1}{ iM\left(\w\right)-\Gamma\ket{e}\bra{e}}+ \frac{1}{-iM\left(\w\right)-\Gamma\ket{e}\bra{e}}\right]\ket{\eta},
\end{split}\ee
with
\be
M\left(\w\right)=\left(\begin{array}{cc}
\w-\Delta - \frac{\Omega_3}{2}  & -\frac{\Omega}{2}\\
-\frac{\Omega}{2} & \w- \Omega_3
\end{array}\right),
\ee
again following the vectorization $\bra{e}=(1,0)$ and $\bra{+}=(0,1)$ as in Eq.(\ref{MatrixM}).

Using the identity
\be
\left(A+B\right)^{-1}=A^{-1}+\frac{1}{1-Tr\{A^{-1}B\}}A^{-1}BA^{-1},
\ee
for an invertible matrix $A$ and a matrix $B$ of rank 1, we obtain
\be
Re\left[S(\w)\right]=\frac{\Gamma\left|\bra{\eta} M\left(\w\right)^{-1}\ket{e}\right|^2}{1+\Gamma^2\left|\bra{e}M\left(\w\right)^{-1}\ket{e}\right|^2}.
\ee
which is the form of Eq.(\ref{realspec}).

\section{Effective coupling of the ground states \label{sec:effective}}

In this section the relationships between the physical Lamb-Dicke parameter $\eta_p$ and the effective $\eta_3$ and between the physical Rabi frequency $\W_p$ and the effective $\W_3$ are derived. The method in \cite{James2007} is followed, where the highly detuned excited level is adiabatically eliminated. Our target Hamiltonian describing the coupling of both ground levels is the following:

\begin{equation}
H_{goal}=\frac{\hbar\Omega_3}{2}\left\lbrace\left[1+i\eta_3\left(b+b^\dagger\right)\right]  \ket{\downarrow} \bra{\uparrow} + H.c.\right\rbrace
\label{Heff}
\end{equation}

The experimental implementation involving a highly detuned Raman coupling is described by the following interaction picture Hamiltonian:

\begin{equation}
\begin{split}
H_{I}=\frac{\Omega_p}{2}&\left\lbrace|e\rangle\langle \uparrow |e^{i\Delta t}\left[1+i\eta_p\left(b^\dagger e^{-i\nu t}+be^{i\nu t}\right)\right] \right. \\
+ &\left.|e\rangle\langle \downarrow |e^{i\Delta t}\left[1-i\eta_p\left(b^\dagger e^{-i\nu t}+be^{i\nu t}\right)\right] +H.c.\right\rbrace.
\end{split}
\label{HI}
\end{equation}

This Hamiltonian consists of the following harmonic terms:
\begin{equation}
\begin{split}
h_1&=\frac{i\Omega_p\eta_p}{2}(|e\rangle\langle \uparrow |-|e\rangle\langle \downarrow |)b^{\dagger}\\
h_2&=\frac{\Omega_p}{2}(|e\rangle\langle \uparrow |+|e\rangle\langle \downarrow |)\\
h_3&=\frac{i\Omega_p \eta_p}{2}(|e\rangle\langle \uparrow |-|e\rangle\langle \downarrow |)b
\end{split}
\end{equation}
with frequency values $\omega_1=\Delta-\nu$, $\omega_2=\Delta$ and $\omega_3=\Delta+\nu$. The derivation of the effective Hamiltonian follows from the formula:
\begin{equation}
H_{eff}(t)=\sum_{m,n=1}^3\overline{\omega_{mn}}^{-1}\left[h_m^{\dagger},h_n\right]\exp [i(\omega_m-\omega_n)t],
\end{equation}
where $\overline{\omega_{mn}}$ is the harmonic average $\overline{\omega_{mn}}^{-1}=1/2(\omega_{m}^{-1}+\omega_{n}^{-1})$. Out of the nine terms originating from this expression, four are of second order in the Lamb-Dicke parameter. The remaining six can be expressed as:
\begin{equation}
\begin{split}
H_{eff}=&\nu a^\dagger a+\frac{\Omega^2_p}{4}\frac{1}{\Delta}(\sigma_z^{(\uparrow,e)}+\sigma_z^{(\downarrow,e)}+\sigma_x^{(\uparrow,\downarrow)})\\
&+\frac{\Omega^2_p\eta_p}{4}\frac{2\Delta^2-\nu^2}{\Delta(\Delta^2-\nu^2)}\hat q \sigma_y^{(\uparrow,\downarrow)}\\
&+\frac{\Omega_p^2\eta_p}{4}\frac{\nu}{\Delta^2-\nu^2}\hat p\sigma_z^{(\uparrow,\downarrow)}.
\end{split}
\end{equation}
where $\hat p= i(b-b^\dagger)$. This Hamiltonian is equivalent to (\ref{Heff}) but for the atomic Stark shifts and the last term, which is proportional to $1/\Delta^2$.

As long as $\Delta\gg\nu$, $H_{eff}$ contains the target interaction $H_{goal}$ with the parametric conditions:
\begin{equation}
\begin{split}
\frac{\Omega_3}{2}&=\frac{\Omega_p^2}{4 \Delta}\\
\frac{\Omega_3\eta_3}{2}&=\frac{\Omega_p^2}{4 \Delta}\eta_p\frac{2\Delta^2-\nu^2}{\Delta^2-\nu^2}
\end{split}
\end{equation}
Solving for the Lamb-Dicke parameter:
\begin{equation}
\eta_3=\eta_p\frac{2\Delta^2-\nu^2}{\Delta^2-\nu^2}
\end{equation}
which yields $\eta_3=2\eta_p$ for the assumed limit $\Delta\gg\nu$.
\end{document}